\documentclass[aps,prl,amsmath,amssymb,tightenlines,epsfig,floatfix,twocolumn,superscriptaddress, longbibliography]{revtex4-1}
\usepackage{graphicx}
\usepackage{dcolumn}	
\usepackage{bm}			
\usepackage{amsfonts}
\usepackage{xspace}
\usepackage{color}
\usepackage{epstopdf}
\usepackage{multirow}

\begin{document}

\newcommand{\psihat}{\ensuremath{\hat{\psi}}\xspace}
\newcommand{\psihatd}{\ensuremath{\hat{\psi}^{\dagger}}\xspace}
\newcommand{\ahat}{\ensuremath{\hat{a}}\xspace}
\newcommand{\Ham}{\ensuremath{\mathcal{H}}\xspace}
\newcommand{\ahatd}{\ensuremath{\hat{a}^{\dagger}}\xspace}
\newcommand{\bhat}{\ensuremath{\hat{b}}\xspace}
\newcommand{\bhatd}{\ensuremath{\hat{b}^{\dagger}}\xspace}
\newcommand{\boldr}{\ensuremath{\mathbf{r}}\xspace}
\newcommand{\dr}{\ensuremath{\,d^3\mathbf{r}}\xspace}
\newcommand{\tr}{\ensuremath{\,\mathrm{Tr}}\xspace}
\newcommand{\dk}{\ensuremath{\,d^3\mathbf{k}}\xspace}
\newcommand{\etal}{\emph{et al.\/}\xspace}
\newcommand{\ie}{i.e.}
\newcommand{\eq}[1]{Eq.~(\ref{#1})\xspace}
\newcommand{\fig}[1]{Fig.~\ref{#1}\xspace}
\newcommand{\abs}[1]{\left| #1 \right|}
\newcommand{\proj}[2]{\left| #1 \rangle\langle #2\right| \xspace}
\newcommand{\Qhat}{\ensuremath{\hat{Q}}\xspace}
\newcommand{\Qhatd}{\ensuremath{\hat{Q}^\dag}\xspace}
\newcommand{\phihatd}{\ensuremath{\hat{\phi}^{\dagger}}\xspace}
\newcommand{\phihat}{\ensuremath{\hat{\phi}}\xspace}
\newcommand{\boldk}{\ensuremath{\mathbf{k}}\xspace}
\newcommand{\boldp}{\ensuremath{\mathbf{p}}\xspace}
\newcommand{\boldsigma}{\ensuremath{\boldsymbol\sigma}\xspace}
\newcommand{\boldalpha}{\ensuremath{\boldsymbol\alpha}\xspace}
\newcommand{\grad}{\ensuremath{\boldsymbol\nabla}\xspace}
\newcommand{\parti}[2]{\frac{ \partial #1}{\partial #2} \xspace}
 \newcommand{\vs}[1]{\ensuremath{\boldsymbol{#1}}\xspace}
\renewcommand{\v}[1]{\ensuremath{\mathbf{#1}}\xspace}
\newcommand{\Psihat}{\ensuremath{\hat{\Psi}}\xspace}
\newcommand{\Psihatd}{\ensuremath{\hat{\Psi}^{\dagger}}\xspace}
\newcommand{\Vhatd}{\ensuremath{\hat{V}^{\dagger}}\xspace}
\newcommand{\Xhat}{\ensuremath{\hat{X}}\xspace}
\newcommand{\Xhatd}{\ensuremath{\hat{X}^{\dag}}\xspace}
\newcommand{\Yhat}{\ensuremath{\hat{Y}}\xspace}
\newcommand{\Jhat}{\ensuremath{\hat{J}}\xspace}
\newcommand{\Yhatd}{\ensuremath{\hat{Y}^{\dag}}\xspace}
\newcommand{\jhat}{\ensuremath{\hat{J}}\xspace}
\newcommand{\lhat}{\ensuremath{\hat{L}}\xspace}
\newcommand{\Nhat}{\ensuremath{\hat{N}}\xspace}
\newcommand{\rhohat}{\ensuremath{\hat{\rho}}\xspace}
\newcommand{\ddt}{\ensuremath{\frac{d}{dt}}\xspace}
\newcommand{\nset}{\ensuremath{n_1, n_2,\dots, n_k}\xspace}
\newcommand{\Var}{\ensuremath{\mathrm{Var}}\xspace}
\newcommand{\notes}[1]{{\color{blue}#1}}

\newcommand{\sss}[1]{{\color{blue}#1}}
\newcommand{\sah}[1]{{\color{magenta}#1}}


\title{High Precision, Quantum-Enhanced Gravimetry with a Bose-Einstein Condensate}
\author{Stuart~S.~Szigeti}
\email{stuart.szigeti@anu.edu.au}
\affiliation{Department of Quantum Science, Research School of Physics, The Australian National University, Canberra 2601, Australia}
\author{Samuel~P.~Nolan}
\affiliation{QSTAR, INO-CNR and LENS, Largo Enrico Fermi 2, Firenze 50125, Italy}
\author{John~D.~Close}
\affiliation{Department of Quantum Science, Research School of Physics, The Australian National University, Canberra 2601, Australia}
\author{Simon A.~Haine}
\affiliation{Department of Quantum Science, Research School of Physics, The Australian National University, Canberra 2601, Australia}

\date{\today}

\begin{abstract}
We show that the inherently large interatomic interactions of a Bose-Einstein condensate (BEC) can enhance the sensitivity of a high precision cold-atom gravimeter beyond the shot-noise limit (SNL). Through detailed numerical simulation, we demonstrate that our scheme produces spin-squeezed states with variances up to 14~dB below the SNL, and that absolute gravimetry measurement sensitivities between two and five times below the SNL are achievable with BECs between $10^4$ and $10^6$ in atom number. Our scheme is robust to phase diffusion, imperfect atom counting, and shot-to-shot variations in atom number and laser intensity. Our proposal is immediately achievable in current laboratories, since it needs only a small modification to existing state-of-the-art experiments and does not require additional guiding potentials or optical cavities.
\end{abstract}

\maketitle

Atom interferometers provide state-of-the-art measurements of gravity~\cite{Peters:1999, Peters:2001, Hu:2013, Hauth:2013, Altin:2013, Freier:2016, Hardman:2016} and gravity gradiometry~\cite{Snadden:1998, Stern:2009, Sorrentino:2014, Biedermann:2015, DAmico:2016, Asenbaum:2017}. Future applications of cold-atom gravimetry are wide ranging~\cite{Bongs:2019}, including inertial navigation~\cite{Jekeli:2005, Battelier:2016, Cheiney:2018}, mineral exploration~\cite{vanLeeuwen:2000,Geiger:2011,Evstifeev:2017}, groundwater monitoring~\cite{Canuel:2018}, satellite gravimetry~\cite{Tino:2013, Carraz:2014}, and weak equivalence principle experiments that test candidate theories of quantum gravity~\cite{Aguilera:2014,Williams:2016,Becker:2018}. These applications require significant improvements to cold-atom gravimeters: improved precision \cite{Dimopoulos:2007}, increased stability~\cite{Menoret:2018}, increased dynamic range~\cite{Lautier:2014}, increased measurement rate~\cite{Rakholia:2014}, and decreased size, weight, and power (SWaP)~\cite{Zoest:2010,Hinton:2017,Wigley:2019}. 

Quantum entanglement offers a promising route to improved cold-atom gravimetry, since it enables relative-phase measurements below the shot-noise limit (SNL). 
Metrologically-useful entanglement has been generated in large cold-atom ensembles via atom-atom~\cite{Esteve:2008, Appel:2009, Lucke:2011, Hamley:2012, Lucke:2014, Muessel:2015, Lange:2018} and atom-light interactions~\cite{Hald:1999, Leroux:2010, Schleier-Smith:2010b, Sewell:2012}, with sub-shot-noise atom interferometry demonstrated in proof-of-principle experiments~\cite{Wasilewski:2010, Riedel:2010, Gross:2010, Leroux:2010b, Schleier-Smith:2010, Hosten:2016, Kruse:2016}. However, no quantum-enhanced (sub-shot-noise) atom interferometer has demonstrated \emph{any} sensitivity to gravity, even in laboratory-based proof-of-principle apparatus. The key challenge is that most methods of generating entangled atomic states are incompatible with the stringent requirements of precision gravimetry. Cold-atom gravimeters require the creation and manipulation of well-defined and well-separated atomic matterwave momentum modes~\cite{Schleich:2013, Kritsotakis:2018}. Although entanglement generation between internal atomic states is relatively mature, no experiment has shown that entanglement between internal states can be converted into entanglement between well-separated, controllable momentum modes suitable for gravimetry. There are promising proposals for creating squeezed momentum states for atom interferometry~\cite{Haine:2013, Szigeti:2014b,Salvi:2018, Shankar:2019}, however these require atom interferometry within an optical cavity which, whilst possible~\cite{Hamilton:2015}, is technically challenging and not always viable (e.g. low-SWaP scenarios). Even if entangled momentum states are available, this does not guarantee that they can be achieved with large atom number sources, nor that a high degree of coherence can be maintained between momentum modes for significant interrogation times. 

In this Letter, we propose a quantum-enhanced ultracold-atom gravimetry scheme that operates in free space. Our scheme uses the large interatomic collisions of a Bose-Einstein condensate (BEC) to generate metrologically-useful entanglement via one-axis twisting (OAT) \cite{Kitagawa:1993, Sorensen:2001}, a nonlinear self-phase modulation that can reduce the relative number fluctuations between two well-defined momentum modes. It does not require additional guiding potentials or optical cavities, making it suitable for low-SWaP scenarios. Our scheme requires only a small modification to existing state-of-the-art experiments, so it is immediately achievable in current laboratories. We show that significant spin squeezing is attainable for large atom numbers and that this spin squeezing results in a useful improvement to absolute gravimetry sensitivity. We further show that our scheme is robust to phase diffusion and common experimental imperfections, including imperfect atom counting and shot-to-shot variations in atom number and laser intensity.

\begin{figure}[t]
	\begin{center}
		\includegraphics[width=0.9\columnwidth]{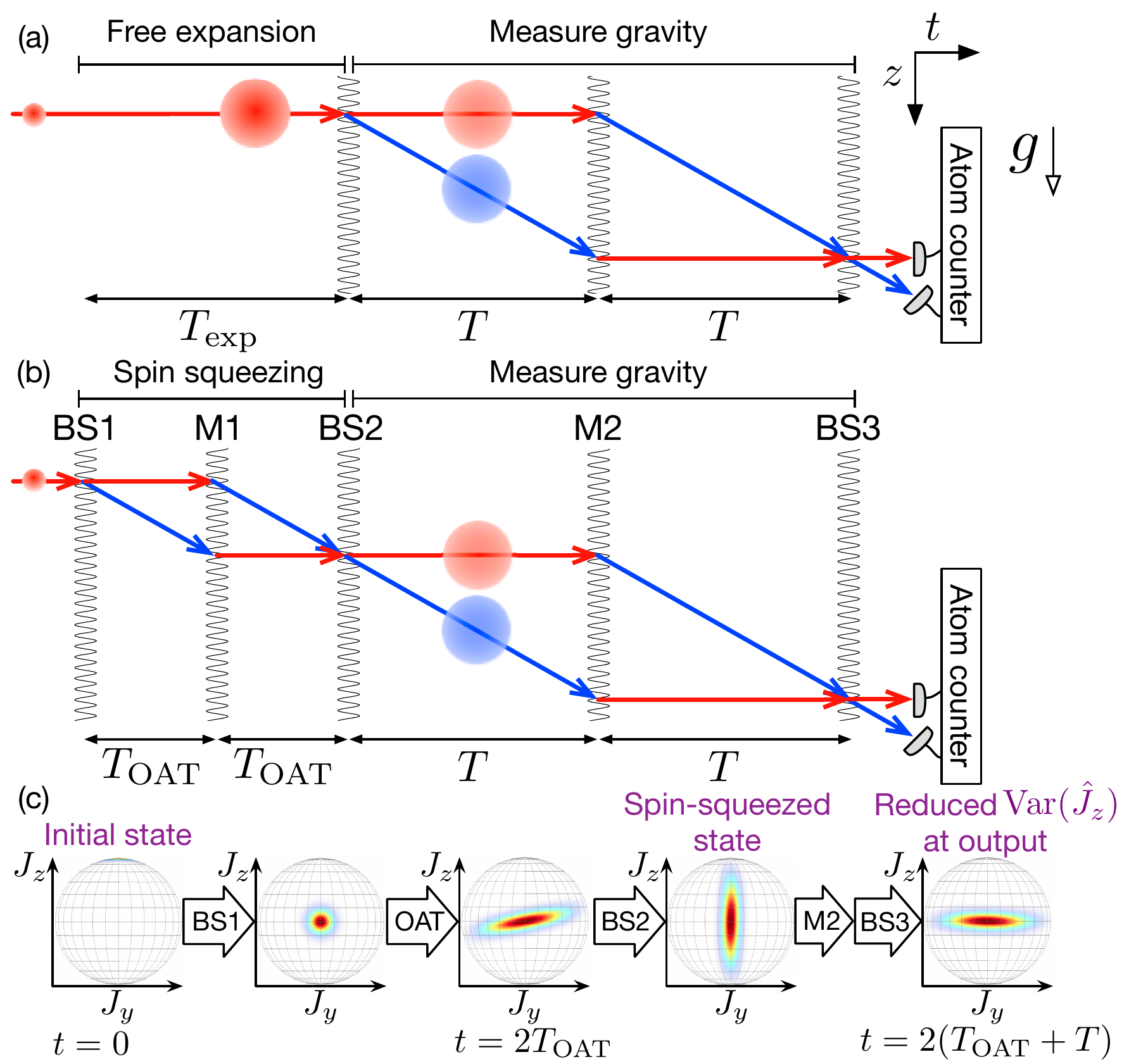}
		\caption{(a) Space-time diagram illustrating SNL gravimetry with a BEC. Unwanted interatomic interactions are reduced by freely expanding the BEC for duration $T_\text{exp}$. A $\pi/2$-$\pi$-$\pi/2$ Raman pulse sequence then creates a MZ interferometer of interrogation time $T$. The two interferometer modes correspond to internal states $|1\rangle$ (red) and $|2\rangle$ (blue) with $\hbar k_0$ momentum separation. (b) Quantum-enhanced ultracold-atom gravimetry. During initial expansion duration $T_\text{exp} = 2 T_\text{OAT}$, the BEC's interatomic interactions generate spin squeezing via OAT. (c) Bloch sphere representation of state during quantum-enhanced gravimetry.}
		\label{fig:scheme}
	\end{center}
\end{figure}

\emph{Gravimetry with a BEC.---} Commonly, an atomic Mach-Zehnder (MZ) is used for gravimetry, where state-changing Raman transitions act as beamsplitters ($\pi/2$ pulses) and mirrors ($\pi$ pulses) \cite{Kasevich:1992}. Raman transitions, achieved with two counter-propagating laser pulses of wavevector $\textbf{k}_L$, coherently couple two internal states $|1\rangle$ and $|2\rangle$. Transitions from $|1\rangle$ to $|2\rangle$ impart $2 \hbar \textbf{k}_L$ momentum to the atoms, giving the momentum separation needed for gravimetry. For $N$ uncorrelated atoms a uniform gravitational acceleration can be measured with single-shot sensitivity $\Delta g =1/(\sqrt{N} k_0 T^2)$, where $k_0$ is the component of $2 \textbf{k}_L$ aligned with gravity and $T$ is the time between pulses (interrogation time) \cite{Kasevich:1992}.

There are advantages to using BECs for precision gravimetry. A BEC's large coherence length and narrow momentum width enables high fringe contrast~\cite{Hardman:2014, Hardman:2016}, improves the efficiency of large momentum transfer beamsplitting~\cite{Debs:2011, Szigeti:2012}, and mitigates many systematic and technical noise effects~\cite{Robins:2013, Abend:2020}. However, a BEC's large interatomic interactions are generally considered an unwanted hinderance. Interatomic collisions couple number fluctuations into phase fluctuations, causing phase diffusion, which degrades sensitivity~\cite{Altin:2011,Altin:2011b}. Consequently, the effects of interatomic collisions are minimized by freely expanding the BEC prior to the MZ's first beamsplitting pulse [Fig~\ref{fig:scheme}(a)], which converts most of the collisional energy to kinetic energy. This reduces phase diffusion and gives excellent mode matching (required for high fringe contrast), since the BEC's spatial mode is largely preserved under free expansion~\cite{Castin:1996,Kagan:1996}.

\emph{Quantum-enhanced gravimetry with a BEC.---} Our scheme, depicted in Fig.~\ref{fig:scheme}(b), is a modification of the standard MZ. Instead of `wasting' the strong interatomic interactions during this initial expansion period, our scheme exploits them with a `state-preparation' interferometer that generates spin squeezing via OAT. Representing the state as a Husimi-$Q$ distribution on the Bloch sphere~\cite{Arecchi:1972, Agarwal:1998}, OAT causes a shearing of the distribution [Fig.~\ref{fig:scheme}(c)]. The second beamsplitter (BS2) rotates the distribution such that it is more sensitive to phase fluctuations within the interferometer, resulting in reduced relative number fluctuations at the output. Necessarily, BS2 is not a 50/50 beamsplitter, with the relative population transfer dependent on the degree of squeezing. Unlike trapped schemes, where interatomic collisions cause unwanted multimode dynamics that make it difficult to match the two modes upon recombination~\cite{Haine:2014}, a BEC's spatial mode is almost perfectly preserved under free expansion, even for large atom numbers and collisional energies. The two modes are therefore well-matched throughout the interferometer sequence. Furthermore, since the collisional energy is converted to kinetic energy during expansion, the interatomic interactions effectively `switch off' after $\sim 10$ms, minimizing their effect during most of the interferometer sequence. For $T \gg T_\text{OAT}$, our scheme enables a gravity measurement with sensitivity (see Supplemental~\cite{supplemental}, which includes Refs.~\cite{Kitagawa:1993,Haine:2005b,Steel:1998,Sinatra:2002,Dennis:2012,Chiofalo:2000,Blakie:2008,Polkovnikov:2010,Opanchuk:2013,Walls:2008,Gardiner:2004b,Olsen:2009,Castin:1996,Altin:2013,Hardman:2016,Sinatra:2011})
\begin{equation}
	\Delta g = \frac{\xi}{\sqrt{N} k_0 T^2} = \frac{1}{\sqrt{N} k_0 T^2}\min_{\theta,\phi}\left(\frac{N \text{Var}(\hat{J}_{\theta,\phi })}{\langle \hat{J}_{\frac{\pi}{2},\phi +\frac{\pi}{2}}\rangle^2}\right)^\frac{1}{2},  \label{Delta_g_squeezed}
\end{equation}
where $\xi \equiv \min_{\theta,\phi} \xi_{\theta,\phi}$ is the spin squeezing parameter~\cite{Wineland:1994,Sorensen:2001}
and $\hat{J}_{\theta,\phi} = \sin \theta \sin \phi \hat{J}_x + \sin \theta \cos \phi \hat{J}_y + \cos \theta \hat{J}_z$. Here $\hat{J}_i = \tfrac{1}{2} \int d \textbf{r} \, \bm{\psi}^\dag(\textbf{r}) \bm{\sigma}_i \bm{\psi}(\textbf{r})$ are pseudospin operators, where $\bm{\sigma}_i$ are the set of Pauli matrices, $\bm{\psi}(\textbf{r}) = ( \hat{\psi}_1(\textbf{r}), \hat{\psi}_2(\textbf{r}) e^{i k_0 z})^T$ with $\hat{\psi}_1(\textbf{r})$ and $\hat{\psi}_2(\textbf{r})$ being field operators describing the BEC's two internal states $|1\rangle$ and $|2\rangle$, respectively, and $i = x,y,z$. Since $[\hat{\psi}_i(\textbf{r}),\hat{\psi}_j^\dag(\textbf{r}')] = \delta_{ij} \delta(\textbf{r}-\textbf{r}')$, $[\hat{J}_i, \hat{J}_j] = i \epsilon_{ijk} \hat{J}_k$ with $\epsilon_{ijk}$ the Levi-Civita symbol. Physically, $\hat{J}_z$ is proportional to the population difference between the two internal states, whilst $\hat{J}_x$ and $\hat{J}_y$ encode coherences between the modes. Equation~(\ref{Delta_g_squeezed}) shows that our scheme is capable of high precision, quantum-enhanced gravimetry provided $\xi < 1$, which is a sufficient condition for spin squeezing~\cite{Pezze:2009}.

\emph{Analytic model of spin squeezing.---} 
In what follows, we assume Raman pulse durations that are much shorter than the timescale for atomic motional dynamics. Typical atom interferometers operate in this regime, allowing us to treat the Raman coupling as an instantaneous beamsplitter unitary $\hat{U}_{\theta,\phi}$ \cite{Kritsotakis:2018}:
\begin{subequations}
\label{BS_psis}
	\begin{align}
		\hat{U}_{\theta,\phi}^\dag\hat{\psi}_1\hat{U}_{\theta,\phi}	&= \cos(\tfrac{\theta}{2}) \hat{\psi}_1 - i e^{i \phi} \sin(\tfrac{\theta}{2}) \hat{\psi}_2 e^{i k_0 z}, \\
		\hat{U}_{\theta,\phi}^\dag\hat{\psi}_2\hat{U}_{\theta,\phi}	&= \cos(\tfrac{\theta}{2}) \hat{\psi}_2 - i e^{-i \phi} \sin(\tfrac{\theta}{2}) \hat{\psi}_1 e^{-i k_0 z},
	\end{align}
\end{subequations}
where $\theta$ and $\phi$ are the beamsplitting angle and phase, respectively. 

Typical spin squeezing models approximate $\hat{\psi}_1(\textbf{r}) \approx u_1(\textbf{r}) \hat{a}_1$ and $\hat{\psi}_2(\textbf{r}) \approx u_2(\textbf{r}) e^{i k_0 z}  \hat{a}_2$, where bosonic modes $\hat{a}_i$ correspond to the two interferometer paths~\cite{Esteve:2008}. This neglects the effect of imperfect spatial-mode overlap on the spin squeezing, which can be substantial \cite{Haine:2014}. Here, we assume $\hat{\psi}_1(\textbf{r},t) = u_1(\textbf{r},t) \hat{a}_1 + \hat{v}_1(\textbf{r},t)$ and $\hat{\psi}_2(\textbf{r},t) = u_2(\textbf{r},t) e^{i k_0 z} \hat{a}_2 + \hat{v}_2(\textbf{r},t)$, where $\int d\textbf{r} \, |u_i(\textbf{r},t)|^2 = 1$ and $\hat{v}_i(\textbf{r},t)$ are `vacuum' operators satisfying $\hat{v}_i(\textbf{r},t) |\Psi\rangle = 0$ and $[ \hat{v}_i(\textbf{r},t), \hat{v}_j^\dag(\textbf{r},t) ] = \delta_{i,j} \left( \delta(\textbf{r}-\textbf{r}') - u_i(\textbf{r},t) u_j^*(\textbf{r}',t)\right)$~\cite{Haine:2005b}.

We calculate $\xi_{\theta,\phi}$ at $t = 2T_\text{OAT}$ immediately before BS2, with the best spin squeezing $\xi$ achieved by optimizing $\theta$ and $\phi$ in the unitary $\hat{U}_{\theta,\phi}$ for BS2. The BEC's evolution between pulses approximately corresponds to OAT Hamiltonian $\hat{H}_\text{OAT}(t) = \hbar \chi(t) \hat{j}_z^2$, where $\hat{j}_z = \frac{1}{2} (\hat{a}_1^\dag \hat{a}_1 - \hat{a}_2^\dag \hat{a}_2)$, $\chi(t) = \chi_{11}(t) + \chi_{22}(t) - 2\chi_{12}(t)$, and $\chi_{ij}(t) = \frac{g_{ij}}{2 \hbar} \int d\textbf{r} \, |u_i(\textbf{r},t)|^2 |u_j(\textbf{r},t)|^2$, with $g_{ij} = 4 \pi \hbar^2 a_{ij}/m$ and $s$-wave scattering lengths $a_{ij}$~\cite{supplemental}.

In the linear squeezing regime, the minimum spin squeezing is~\cite{supplemental}
\begin{equation}
	\xi^2 \approx \frac{1 - \frac{1}{2}|\mathcal{Q}| N \lambda (\sqrt{4 + |\mathcal{Q}|^2 N^2 \lambda^2} - |\mathcal{Q}| N \lambda)}{|\mathcal{Q}|^2}, \label{xi_min}
\end{equation}
where $\lambda \equiv \int_0^{2 T_\text{OAT}} dt' \chi(t')$ and $\mathcal{Q} \equiv |\mathcal{Q}| e^{i \varphi} = \int d \textbf{r} u_1^*(\textbf{r}, 2T_\text{OAT}) u_2(\textbf{r}, 2T_\text{OAT})$.  Physically, $|\mathcal{Q}|$ quantifies how well the interferometer modes $\hat{a}_1$ and $\hat{a}_2$ are spatially matched at BS2 ($t = 2 T_\text{OAT}$), with $|\mathcal{Q}| = 1$ indicating perfect spatial overlap. Minimum spin squeezing requires $\theta \approx \tfrac{3\pi}{2} - \frac{1}{2} \tan^{-1} \left[2/(N |\mathcal{Q}| \lambda) \right]$ and $\phi = -\varphi$ for the BS2 unitary. 
Since $\lambda > 0$, Eq.~(\ref{xi_min}) shows that $\xi < 1$ always, provided good mode overlap $|\mathcal{Q}|$ is maintained.

\begin{figure}[t]
	\begin{center}
		\includegraphics[width=0.9\columnwidth]{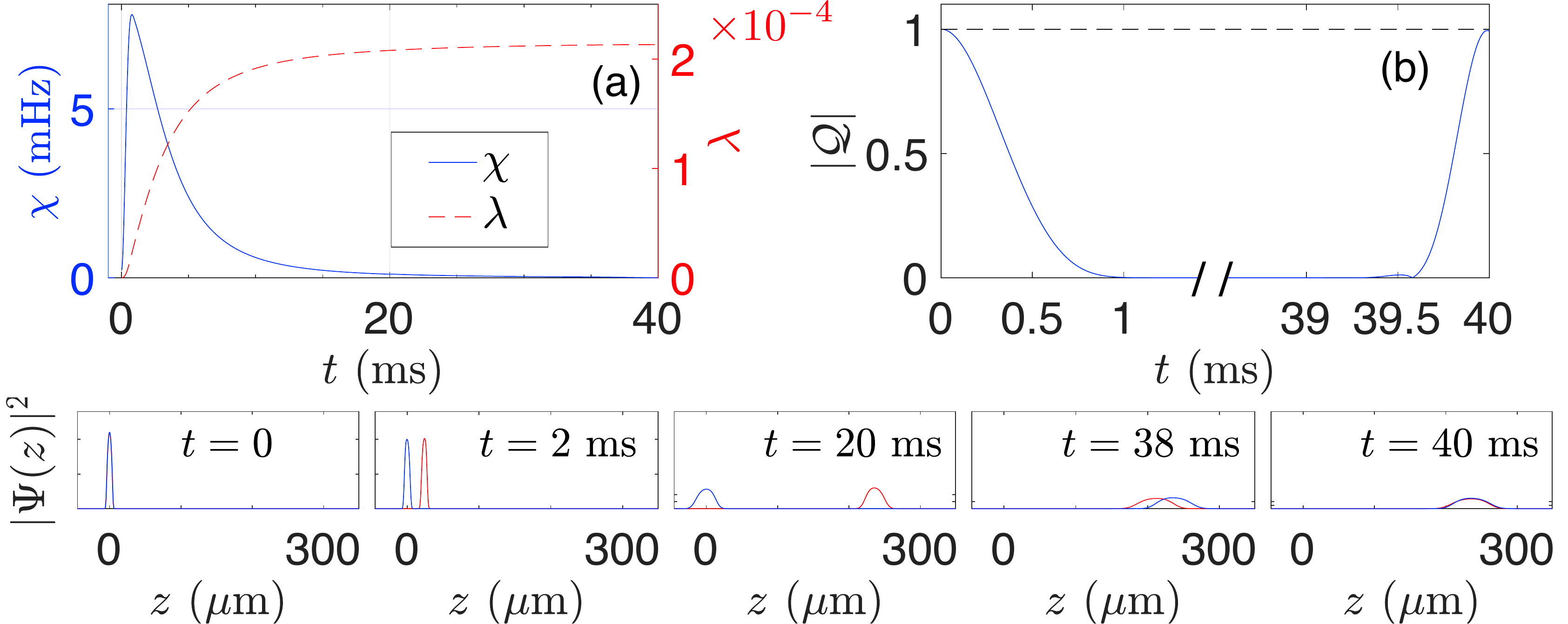}
		\caption{Analytic spin squeezing model parameters determined from a GPE simulation of our scheme up to $t=2T_\text{OAT}$, with $T_\text{OAT} = 20$ms and an $N = 10^4$ atom BEC initially prepared in a spherical harmonic trap of frequency 50Hz. (a) Effective squeezing rate $\chi(t)$ (blue, solid) and squeezing degree $\lambda(t)$ (orange, dashed). (b) Mode overlap $|\mathcal{Q}(t)|$. (Bottom) Normalized density slices at radial coordinate $r=0$.}
		\label{fig:analytics}
	\end{center}
\end{figure}

We estimate $\mathcal{Q}$ and $\lambda$ by numerically solving the two-component Gross-Pitaevskii equation (GPE) for mean-field wavefunctions $\Psi_i(\textbf{r},t)$ and identifying $u_i(\textbf{r},t) = \Psi_i(\textbf{r},t)/\sqrt{N}$~\cite{supplemental}. For concreteness, we take $|1\rangle$ and $|2 \rangle$ as the $|F=1,m_F=0\rangle$ and $|F=2,m_F=0\rangle$ hyperfine states, respectively, of $^{87}$Rb with $(a_{11}, a_{22}, a_{12}) = (100.4, 95.0,97.66) a_0$ and $k_0 = 2 k_L = 1.61 \times 10^7$m$^{-1}$ ($780$nm D2 transition). Figure~\ref{fig:analytics} illustrates the key advantages of our scheme by plotting how $\chi(t)$, $\lambda(t) =  \int_0^t dt' \chi(t')$ and $|\mathcal{Q}(t)| = |\int d \textbf{r} u_1^*(\textbf{r}, t) u_2(\textbf{r},t)|$ vary during the interferometer sequence. All three scattering lengths are of similar magnitude, so during the short duration where the two modes are strongly overlapped, $\chi(t)$ is almost zero and little spin squeezing is produced. However, the two modes rapidly separate ($\sim 1$ms) whilst the interatomic interactions are still significant, substantially increasing $\lambda(t)$. Most of this increase occurs over the next 10ms; after this, free expansion rapidly reduces the collisional energy and therefore $\chi(t)$. Fortunately, this expansion is self-similar, largely preserving the mode shape, allowing high spatial-mode overlap ($|\mathcal{Q}|\sim 1$) at the interferometer output.

\emph{Spin squeezing results.---} Although this analytic model provides qualitative insights into our scheme's viability, quantitative modelling requires a multimode description that, unlike the GPE, incorporates the effect of quantum fluctuations. This description is provided by the truncated Wigner (TW) method, which has successfully modelled BEC dynamics in regimes where nonclassical particle correlations become important~\cite{Steel:1998, Sinatra:2002, Norrie:2006, Opanchuk:2012, Drummond:2017, Johnson:2017, Szigeti:2017, Brown:2018, Haine:2018}. In this approach, the BEC dynamics are efficiently simulated by a set of stochastic differential equations (SDEs), with averages over the solutions of these SDEs corresponding to symmetically-ordered operator expectations~\cite{supplemental}. 

\begin{figure}[t!]
	\begin{center}
		\includegraphics[width=\columnwidth]{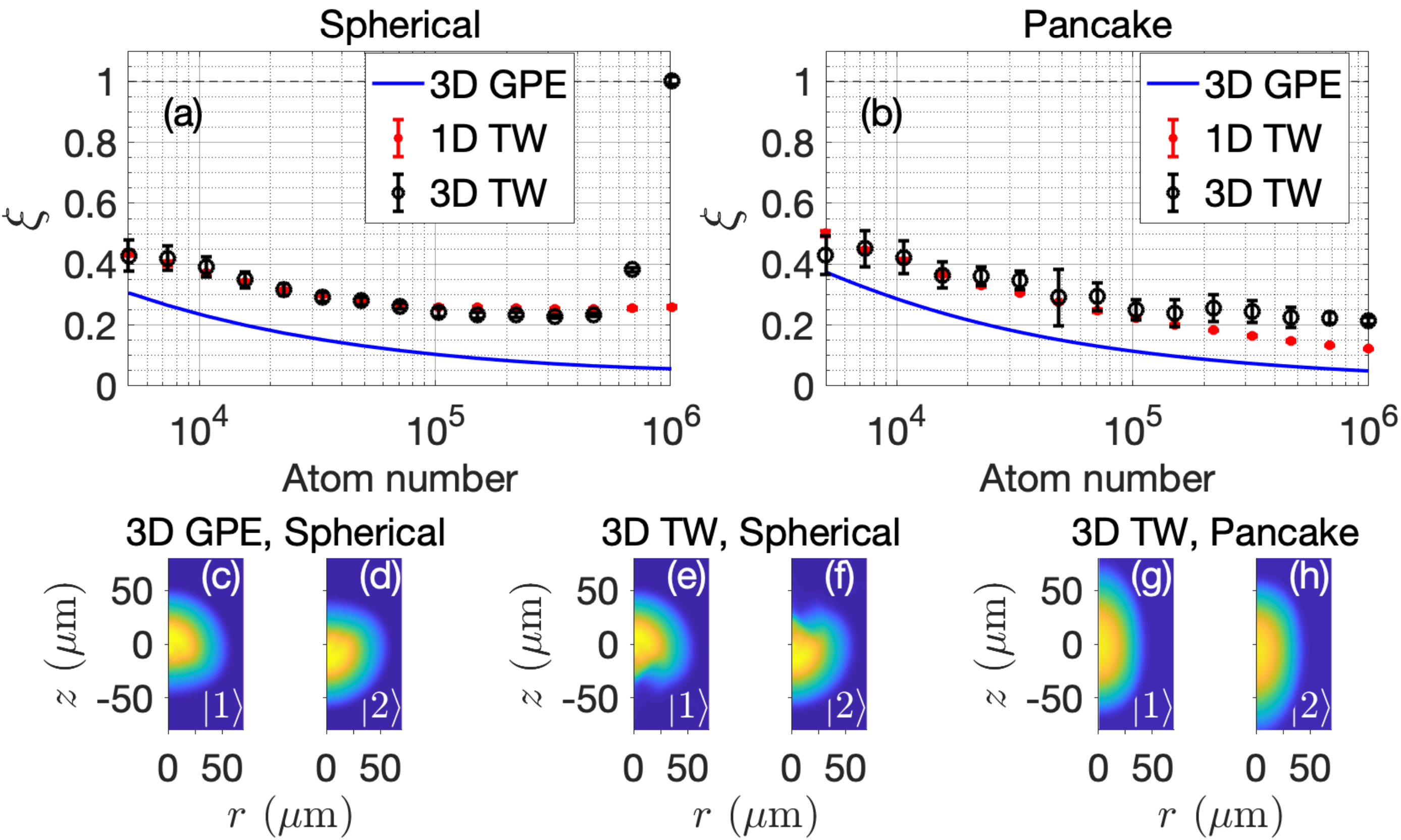}
		\caption{Minimum spin squeezing parameter $\xi$ for $T_\text{OAT} = 10$ms and atom number $N$~\cite{supplemental}. In (a) the BEC is initially prepared in a spherical harmonic trap ($f_r = f_z = 50$Hz), whereas in (b) an initial `pancake' BEC is prepared in a cylindrically-symmetric harmonic trap ($f_r = 32$Hz, $f_z = 160$Hz). TW simulations are compared to Eq.~(\ref{xi_min}) with model parameters determined from GPE simulations (`3D GPE'). (c)-(h) Density profiles for $N=10^6$ at $t=2T_\text{OAT}$. The analytic model fails here for the spherical BEC case since spontaneous scattering degrades mode overlap.}
		\label{fig:SSplot}
	\end{center}
\end{figure}

Figure~\ref{fig:SSplot} compares the spin squeezing parameter computed from our analytic model Eq.~(\ref{xi_min}), with $\lambda$ and $|\mathcal{Q}|$ determined from 3D GPE simulations, to a direct computation of $\xi$ via 3D TW simulations. We consider two scenarios: an initial spherical BEC prepared in a spherical harmonic trap of frequency 50 Hz [Fig.~\ref{fig:SSplot}(a)] and an initial `pancake' BEC prepared in a cylindrically-symmetric harmonic trap with frequencies $(f_r,f_z) = (32,160)$Hz in the radial and $z$ directions [Fig.~\ref{fig:SSplot}(b)]. Although the analytic model correctly captures the atom-number dependence, it overestimates the degree of squeezing by roughly a factor of two. An exception is for the largest atom numbers considered in the spherical case, where TW predicts much worse squeezing. For these atom numbers, the interatomic interactions are sufficiently strong such that intercomponent scattering strongly degrades the mode overlap, even though the clouds are initially overlapped for only $\sim 1$ms  [Figs.~\ref{fig:SSplot}(e) and (f)]. This is not seen in the GPE simulations [Figs.~\ref{fig:SSplot}(c) and (d)] which neglect spontaneous scattering processes that clearly matter. In contrast, for an initially pancake-shaped BEC that is spatially tight in $z$, the two modes spatially separate on a timescale much faster than the spherical case. This mitigates the effect interatomic interactions have on mode matching [Figs.~\ref{fig:SSplot}(g) and (h)], allowing significant squeezing even for $N = 10^6$ atoms.

\emph{Simulation of full interferometer sequence.---} Although the spin squeezing parameter shows that our scheme produces significant spin squeezing, it does not confirm that this spin squeezing leads to a more sensitive measurement of $g$. Residual interatomic interactions may further degrade mode overlap during the remainder of the interferometer sequence and can couple to quantum fluctuations in $\hat{J}_z$, causing phase diffusion~\cite{Altin:2011,Altin:2011b}. Both effects may degrade the sensitivity from the value predicted by Eq.~(\ref{Delta_g_squeezed}). We confirm that these effects are not significant in our scheme by simulating the full interferometer sequence and directly computing the sensitivity via $\Delta g^2 = \text{Var}(\hat{J}_z) / (\partial \langle \hat{J}_z \rangle / \partial g )^2$. 3D TW simulations of the full interferometer sequence are computationally infeasible, since they require prohibitively large grids and numbers of trajectories. 
Instead, we use an effective 1D TW model for these simulations, which assumes a Thomas-Fermi radial profile that self-similarly expands according to scaling solutions~\cite{supplemental}. As shown in Fig.~\ref{fig:SSplot}, this model perfectly agrees with 3D TW simulations except for the largest atom numbers.

\begin{figure}[t]
	\begin{center}
		\includegraphics[width=0.9\columnwidth]{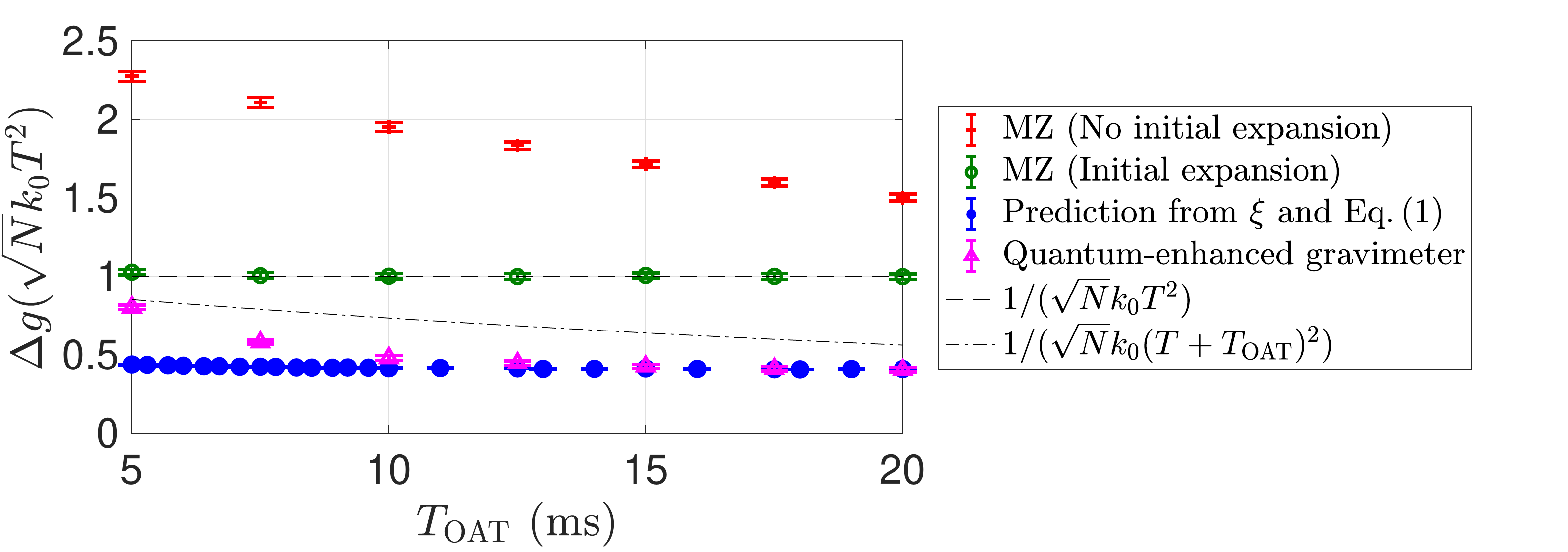}
		\caption{1D TW calculations of sensitivity $\Delta g$ for an $N = 10^4$ atom BEC initially prepared in a cylindrically-symmetric harmonic trap ($f_r = 32$Hz, $f_z = 160$Hz). From top to bottom: (red) MZ with total interrogation time $T + T_\text{OAT}$ (no initial period of free expansion), (green) BEC undergoes free expansion for duration $2 T_\text{OAT}$, followed by MZ of interrogation time $T$ [Fig.~\ref{fig:scheme}(a)]; (magenta) quantum-enhanced BEC gravimetry [Fig.~\ref{fig:scheme}(b)]; (blue) Eq.~(\ref{Delta_g_squeezed}) with $\xi$ computed via TW. All four cases have the same total duration $2(T_\text{OAT} + T)$ with $T = 60$ms. The SNL for an ideal MZ of interrogation time $T$ (dashed) and $T+T_\text{OAT}$ (dot-dashed) are marked for comparison. Our quantum-enhanced scheme always outperforms MZ schemes, even when phase diffusion is non-negligible.}
		\label{fig:phase_diffusion}
	\end{center}
\end{figure}

Our scheme's sensitivity for an initial pancake BEC of $N = 10^4$ atoms and $T = 60$ms is shown in Fig.~\ref{fig:phase_diffusion}. Although phase diffusion degrades the sensitivity for small $T_\text{OAT}$, its effect rapidly reduces for increasing $T_\text{OAT}$, becoming negligible for $T_\text{OAT} \gtrsim 15$ms. We compare our scheme to two SNL cold-atom gravimeters with the same initial BEC and total interferometer time $2(T_\text{OAT} + T)$: (1) the conventional BEC gravimeter depicted in Fig.~\ref{fig:scheme}(a) (MZ with initial $T_\text{exp} = 2 T_\text{OAT}$ period of free expansion) and (2) a MZ with no initial period of free expansion, thereby having an increased interrogation time $T + T_\text{OAT}$. As expected, the former has negligible phase diffusion, attaining the ideal SNL result $\Delta g = 1 / (\sqrt{N} k_0 T^2)$. The latter suffers from considerable phase diffusion, far outweighing the benefit of increased interrogation time. Our scheme outperforms both SNL gravimeters, demonstrating the clear benefit of using the initial $2 T_\text{OAT}$ period to produce spin squeezing.

\emph{Experimental imperfections.---} Finally, we assess the effect of three common experimental imperfections. 

\begin{figure}[t!]
	\begin{center}
		\includegraphics[width=\columnwidth]{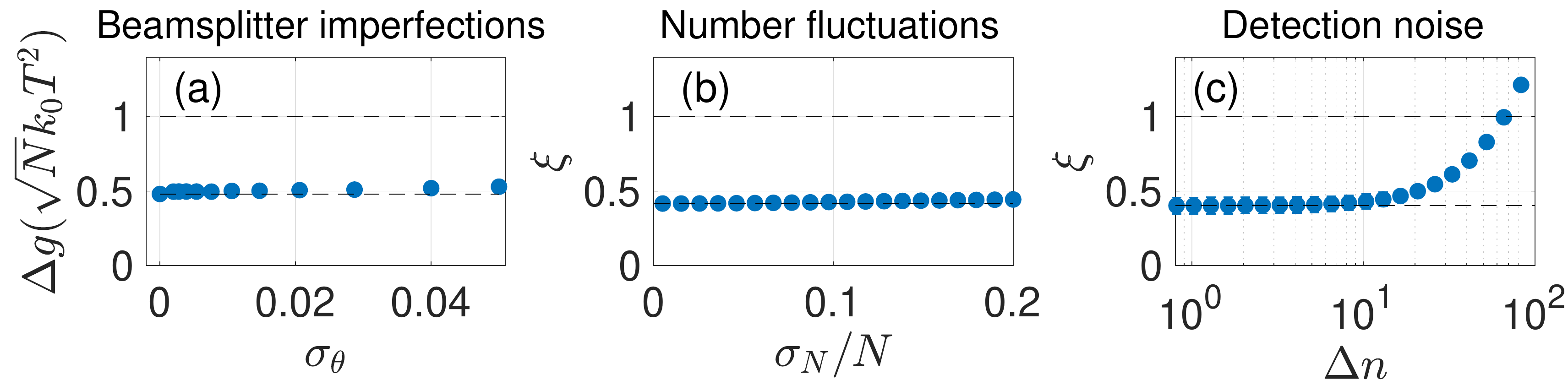}
		\caption{The effect on our scheme of (a) Gaussian shot-to-shot beamsplitting angle fluctuations of variance $\sigma_\theta^2$, (b) Gaussian shot-to-shot atom-number fluctuations of variance $\sigma_N^2$, and (c) imperfect atom detection of resolution $\Delta n$. Here $N=10^4$, $T_\text{OAT} = 10$ms, with an initial BEC prepared in a cylindrically-symmetric harmonic trap ($f_r = 32$Hz, $f_z = 160$Hz). In (a) the sensitivity was obtained via 1D TW simulations of the full interferometer ($T = 60$ms), whereas (b) and (c) computed the spin squeezing parameter from 3D TW simulations.}
		\label{fig:imperfections}
	\end{center}
\end{figure}

\emph{(i) Shot-to-shot fluctuations in laser intensity.---} Although the laser pulse intensity is stable during a single interferometer run, it can vary between experimental runs~\cite{Karcher:2018}. Such shot-to-shot intensity fluctuations cause an offset $\delta \theta$ to the angle of all beamsplitters and mirrors in that run, where $\delta \theta$ varies from shot-to-shot~\cite{Hosten:2016}. To first order, $\delta \theta \approx 2 \Delta f$, where $\Delta f$ is the fractional change in the population ratio due to imperfect beamsplitting (e.g. $\Delta f = 0.02$ means that a 50/50 beamsplitter is instead performed as a 48/52 beamsplitter). We simulated the full interferometer sequence assuming that all five laser pulses suffered from Gaussian-distributed shot-to-shot fluctuations $\delta \theta$ of variance $\sigma_\theta^2$. As shown in Fig.~\ref{fig:imperfections}(a), these shot-to-shot fluctuations have a relatively small effect on $\Delta g$, since common rotation errors from the different pulses largely cancel. 

\emph{(ii) Shot-to-shot fluctuations in atom number.---} The optimal rotation angle $\theta$ for BS2 depends on the atom number. This cannot be known precisely and varies 10-20\% for different experimental runs~\cite{Hardman:2014,Hardman:2016}. Consequently, $\theta$ will deviate from the optimum from shot-to-shot, degrading $\xi$. We quantify this by assuming Gaussian-distributed shot-to-shot atom number fluctuations about mean $N$ with variance $\sigma_N^2$. To leading order, optimal BS2 parameters for atom number $N$ give $\xi(\sigma_N) \lesssim \xi + \tfrac{1}{2 |\mathcal{Q}|^2}(\sigma_N / N)^2$~\cite{supplemental}, so shot-to-shot atom number fluctuations weakly impact the spin squeezing. This is confirmed by TW simulations [Fig.~\ref{fig:imperfections}(b)].

\emph{(iii) Imperfect atom detection.---} We model imperfect detection resolution as a Gaussian noise of variance $(\Delta n)^2$, corresponding to uncertainty $\Delta n$ in the measured atom number. Imperfect detection increases the variance in $\hat{J}_z$, giving poorer sensitivity $\Delta g^2 = (\text{Var}(\hat{J}_z) + \Delta j_z^2) / (\partial \langle \hat{J}_z \rangle / \partial g )^2$, where $\Delta j_z = \Delta n/\sqrt{2}$. 
Then $\Delta g$ is given by Eq.~(\ref{Delta_g_squeezed}) with a modified spin squeezing parameter $\xi({\Delta n})^2 \approx \xi^2 + (2/N) \Delta n^2$ \cite{resolutionfootnote}. 
Figure~\ref{fig:imperfections}(a) plots the dependence of $\xi$ on $\Delta n$. Although the requirements are stringent, they are achievable and comparable to other spin-squeezing experiments. For example, Ref.~\cite{Engelsen:2017} reports $\Delta n \sim 8$ for an $N = 5\times 10^5$ atom ensemble, which would minimally impact our scheme's sensitivity.

\emph{Conclusions.---} We have presented a scheme for quantum-enhanced gravimetry that exploits a BEC's inherently strong interatomic interactions, rather than simply removing them through an initial free expansion period. This scheme allows high-precision gravimetry up to a factor of five below the SNL and is robust to a range of experimental imperfections. Concretely, a quantum-enhanced gravimeter with $N = 10^6$ and $\xi = 0.2$ is equivalent to a SNL gravimeter with $N = 2.5 \times 10^7$ -- a challenging atom number to attain with current cooling methods~\cite{Robins:2013}. Equivalently, for a fixed sensitivity, $\xi = 0.2$ allows a five-fold reduction in device size, enabling the more compact gravimeters needed for low-SWaP scenarios. Larger values of $k_0$, obtainable via Bragg pulses~\cite{Altin:2013}, could reduce the initial period of time where the two modes are overlapping. This would further reduce the deleterious effect of spontaneous scattering at large $N$, potentially allowing more significant degrees of spin squeezing. Since our proposal operates in free space, requiring only a small modification to existing laboratory setups, it provides a path towards realizing quantum-enhanced cold-atom gravimetry in the immediate future.

\begin{acknowledgements}
\emph{Acknowledgements.---} We acknowledge fruitful discussions with Chris Freier, Kyle Hardman, Joseph Hope, Nicholas Robins, and Paul Wigley. This project was partially funded by a Defence Science and Technology Group Competitive Evaluation Research Agreement, Project MyIP: 7333. SSS was supported by an Australian Research Council Discovery Early Career Researcher Award (DECRA), project DE200100495. SPN acknowledges funding from the H2020 QuantERA ERA-NET Cofund in Quantum Technologies, project CEBBEC. This research was undertaken with the assistance of resources and services from the National Computational Infrastructure (NCI), which is supported by the Australian Government.
\end{acknowledgements}

\bibliography{OAT_bib.bib}

\begin{widetext}

\renewcommand{\thefigure}{[S\arabic{figure}]} 
\renewcommand{\theequation}{S\arabic{equation}} 
\renewcommand\labelenumi{[\theenumi]}
\setcounter{equation}{0}
\setcounter{figure}{0}

\section{Supplemental Material: High Precision, Quantum-Enhanced Gravimetry with a Bose-Einstein Condensate}

In this supplemental material we provide (1) a derivation of the gravitational sensitivity (Eq.~(1) of the main text), (2) details of our analytic model of spin squeezing, culminating in a derivation of Eq.~(3) of the main text, (3) details of our numerical Gross-Pitaevskii equation (GPE) simulations, (4) details of our 3D truncated Wigner (TW) simulations, (5) a derivation of our effective 1D TW simulation model and details of our full interferometer simulations using this effective model, (6) justification for our use of zero-temperature models in our analysis, (7) plots of the optimal second beamsplitting parameters used to attain the minimum spin squeezing reported in Fig.~3 of the main text, with a brief description of how to optimize these parameters in an experiment, and (8) an analytic calculation showing that shot-to-shot fluctuations in the total atom number only weakly degrade the spin squeezing parameter.

\section{Derivation of gravitational sensitivity, Eq.~(1)}
Here we show that our scheme can be used to measure gravity at a sensitivity given by Eq.~(1) of the main text. Our derivation uses the pseudospin operators
\begin{subequations} \label{pseudo_spin_operators}
\begin{align}
	\hat{J}_x	&= \frac{1}{2}\int d\textbf{r} \left( \hat{\psi}_1^\dag(\textbf{r})\hat{\psi}_2(\textbf{r}) e^{-i k_0 z} +  \hat{\psi}_1(\textbf{r})\hat{\psi}_2^\dag(\textbf{r}) e^{i k_0 z}\right), \\
	\hat{J}_y	&= -\frac{i}{2}\int d\textbf{r} \left( \hat{\psi}_1^\dag(\textbf{r})\hat{\psi}_2(\textbf{r}) e^{-i k_0 z} -  \hat{\psi}_1(\textbf{r})\hat{\psi}_2^\dag(\textbf{r}) e^{i k_0 z}\right), \\
	\hat{J}_z	&= \frac{1}{2}\int d\textbf{r} \left( \hat{\psi}_1^\dag(\textbf{r})\hat{\psi}_1(\textbf{r}) -  \hat{\psi}_2^\dag(\textbf{r})\hat{\psi}_2(\textbf{r}) \right),
\end{align}
\end{subequations}
which satisfy $[\hat{J}_i, \hat{J}_j] = i \epsilon_{ijk} \hat{J}_k$, where $\epsilon_{ijk}$ is the Levi-Civita symbol. The beamsplitter unitary $\hat{U}_{\theta,\phi}$, defined by Eqs~(2) of the main text, transforms the pseudospin operators as
\begin{subequations}
\begin{align}
	\hat{U}_{\theta,\phi}^\dag \hat{J}_x \hat{U}_{\theta,\phi}	&= \cos \phi \left( \cos \phi \hat{J}_x - \sin \phi \hat{J}_y \right) + \cos \theta \sin \phi \left( \sin \phi \hat{J}_x + \cos \phi \hat{J}_y \right) - \sin \theta \sin \phi \hat{J}_z, \\
	\hat{U}_{\theta,\phi}^\dag \hat{J}_y \hat{U}_{\theta,\phi}	&= -\sin \phi \left( \cos \phi \hat{J}_x - \sin \phi \hat{J}_y \right) + \cos \theta \cos \phi \left( \sin \phi \hat{J}_x + \cos \phi \hat{J}_y \right) - \sin \theta \cos \phi \hat{J}_z, \\
	\hat{U}_{\theta,\phi}^\dag \hat{J}_z \hat{U}_{\theta,\phi}	&= \sin \theta \left( \sin \phi \hat{J}_x + \cos \phi \hat{J}_y \right) + \cos \theta \hat{J}_z.
\end{align}
\end{subequations}
Geometrically, the beamsplitting operation corresponds to a rotation of the spin vector $\hat{\textbf{J}} = ( \hat{J}_x, \hat{J}_y, \hat{J}_z )$ by some angle $\alpha$ about some axis $\textbf{s}$: $\hat{U}_{\theta,\phi} = \exp[-i \alpha(\theta,\phi) \hat{\textbf{J}} \cdot \textbf{s}(\theta,\phi)]$. In particular, $\hat{U}_{\theta,0} = \exp(-i \theta \hat{J}_x)$ (i.e. a rotation about the $J_x$-axis by angle $\theta$) and $\hat{U}_{\theta,-\pi/2} = \exp(-i \theta \hat{J}_y)$ (i.e. a rotation about the $J_y$-axis by angle $\theta$).

We model the effect of a uniform gravitational acceleration $g$ over interrogation period $T$ as a relative phase shift $\varphi = g k_0 T^2$ between the modes: $\hat{\psi}_1(\textbf{r}) \to \hat{\psi}_1(\textbf{r}) \exp(- i \varphi / 2), \hat{\psi}_2(\textbf{r}) \to \hat{\psi}_2(\textbf{r}) \exp(i \varphi / 2)$. In terms of the pseudospin operators, this corresponds to a rotation about the $J_z$-axis by angle $\varphi$:
\begin{subequations}
\begin{align}
	e^{i \varphi \hat{J}_z} \hat{J}_x e^{-i \varphi \hat{J}_z}	&= \cos \varphi \hat{J}_x - \sin \varphi \hat{J}_y, \\
	e^{i \varphi \hat{J}_z} \hat{J}_y e^{-i \varphi \hat{J}_z}	&= \sin\varphi \hat{J}_x + \cos \varphi \hat{J}_y \\
	e^{i \varphi \hat{J}_z} \hat{J}_z e^{-i \varphi \hat{J}_z}	&= \hat{J}_z.
\end{align}
\end{subequations}

In the freely-falling frame, the BEC's evolution between beamsplitting pulses is given by the Hamiltonian 
\begin{equation}
	\hat{H} = \sum_{i=1,2} \int d\textbf{r} \, \hat{\psi}_i^\dag(\textbf{r})\left(-\frac{\hbar^2}{2m}\nabla^2 \right)\hat{\psi}_i(\textbf{r}) + \sum_{i,j=1,2} \frac{g_{ij}}{2} \int d\textbf{r} \, \hat{\psi}_i^\dag(\textbf{r})\hat{\psi}_j^\dag(\textbf{r})\hat{\psi}_j(\textbf{r})\hat{\psi}_i(\textbf{r}), \label{full_field_Hamiltonian}
\end{equation}
where $g_{ij} = 4 \pi \hbar^2 a_{ij}/m$ for $s$-wave scattering lengths $a_{ij}$. The unitary corresponding to this evolution is: $\hat{U}_\text{evo}(t,t_0) = \exp[-\frac{i}{\hbar} \hat{H}(t-t_0)]$. This operation causes both self-similar expansion of the BEC and the spin squeezing in our scheme. This is shown more clearly below in subsection `One-axis twisting due to interatomic interactions', however for now it is sufficient to keep this unitary general.

Our scheme, depicted in Fig.~1(b) of the main text, begins with all $N$ atoms in internal state $|1\rangle$ (i.e. a $\hat{J}_z$ eigenstate at the top of the Bloch sphere). The operations defining our scheme are:
\begin{enumerate}
	\item {[BS1]} A $\pi/2$-pulse coherently prepares a 50/50 superposition of atoms in state $|1\rangle$ (momentum $\textbf{p}$) and state $|2\rangle$ (momentum $\textbf{p} + \hbar k_0 \hat{\textbf{z}}$). This is a $\pi/2$ rotation about the $J_y$-axis ($\hat{U}_{\pi/2,-\pi/2}$) and results in a coherent spin state polarized along the $J_x$-axis.
	\item The BEC evolves according to $\hat{U}_\text{evo}(T_\text{OAT},0)$.
	\item {[M1]} A $\pi$-pulse coherently reflects the two atomic matterwaves at $t = T_\text{OAT}$ ($\pi$ rotation about $J_x$-axis: $\hat{U}_{\pi,0}$).
	\item The BEC evolves according to $\hat{U}_\text{evo}(2T_\text{OAT},T_\text{OAT})$. At $t = 2 T_\text{OAT}$ the self-similar expansion of the BEC means that the interatomic interactions are negligible. Therefore, for $t > 2 T_\text{OAT}$, $\hat{U}_\text{evo}$ only causes a propagation phase shift which cancels at the final beamsplitter. 
	\item We include the effect of gravity from $t = 0$ to $t = 2 T_\text{OAT}$ by applying a $\phi_1 = k_0 g T_\text{OAT}^2$ rotation about the $J_z$-axis: $\exp(-i \phi_1 \hat{J}_z)$. 
	\item {[BS2]} At time $t = 2 T_\text{OAT}$, the second beamsplitter $\hat{U}_{\theta ,\phi}$ prepares a phase sensitive state (i.e. a state with minimum variance in $\hat{J}_y$). 
	\item {[M2]} A second $\pi$-pulse is applied at $t = 2T_\text{OAT} + T$ ($\pi$ rotation about $J_x$-axis: $\hat{U}_{\pi,0}$).
	\item We include the effect of gravity from $t = 2T_\text{OAT}$ to $t = 2 T_\text{OAT} + 2T$ by applying a $\phi_2 = k_0 g T^2$ rotation about the $J_z$-axis: $\exp(-i \phi_2 \hat{J}_z)$. 
	\item {[BS3]} At $t = 2 T_\text{OAT} + 2T$, when the two matterwaves are spatially overlapping, a  final $\pi/2$ pulse with phase $\phi_\text{BS3}$ ($-\pi/2$ rotation about $J_x$: $\hat{U}_{-\pi/2,\phi_\text{BS3}}$] recombines the two matterwaves. As shown below, the additional phase shift $\phi_\text{BS3}$ adjusts for the gravitational phase shifts $\phi_1$ and $\phi_2$ and BS2 phase shift $\phi$. 
\end{enumerate}
The overall unitary describing this scheme is therefore
\begin{equation}
	\hat{U}_\text{total} = \hat{U}_\text{linear}\hat{U}_\text{nonlinear},
\end{equation}
where
\begin{subequations}
\begin{align}
	\hat{U}_\text{nonlinear} 	&=\hat{U}_\text{evo}(2T_\text{OAT},T_\text{OAT})\hat{U}_{\pi,0}\hat{U}_\text{evo}(T_\text{OAT},0)\hat{U}_{\pi/2,-\pi/2}, \\
	\hat{U}_\text{linear} 	&=  \hat{U}_{-\pi/2,\phi_\text{BS3}}e^{-i \phi_2 \hat{J}_z}\hat{U}_{\pi,0}\hat{U}_{\theta ,\phi} e^{-i \phi_1 \hat{J}_z}.
\end{align}
\end{subequations}
Here $\hat{U}_\text{nonlinear}$ includes operations 1 - 4 above, which is independent of $g$ and is nonlinear in the pseudospin operators. In contrast, the unitary $\hat{U}_\text{linear}$ includes operations 5 - 9 above, and consists entirely of linear rotations. This allows us to derive an expression for the sensitivity with respect to operator expectations taken at time $t = 2T_\text{OAT}$ (immediately before BS2). 

Assuming a number-difference ($\hat{J}_z$) measurement at the interferometer output, the sensitivity to a gravitational acceleration is given by the usual linear error propagation formula:
\begin{equation}
	\Delta g = \frac{\sqrt{\text{Var}(\hat{J}_z^\text{out})}}{|\partial \langle \hat{J}_z^\text{out} \rangle / \partial g |}, \label{sensitivity_formula}
\end{equation}
where $\hat{J}_i^\text{out} \equiv \hat{U}_\text{total}^\dag \hat{J}_i \hat{U}_\text{total}$ for $i = x,y,z$. Denoting $\hat{J}_i^\text{OAT} \equiv \hat{U}_\text{nonlinear}^\dag \hat{J}_i \hat{U}_\text{nonlinear}$, it is not too difficult to show that
\begin{align}
	\hat{J}_z^\text{out}	&= C_x \hat{J}_x^\text{OAT} + C_y \hat{J}_y^\text{OAT} + C_z \hat{J}_z^\text{OAT}, \label{Jz_out_exp}
\end{align}
where
\begin{subequations}
\label{Jz_out_coeffs}
\begin{align}
	C_x 	&= \sin(\phi_2 + \phi + \phi_\text{BS3})\cos(\phi_1+\phi) + \cos(\phi_2 + \phi + \phi_\text{BS3})\sin(\phi_1+\phi)\cos \theta , \\ 
	C_y	&= \sin(\phi_2 + \phi + \phi_\text{BS3})\sin(\phi_1+\phi) + \cos(\phi_2 + \phi + \phi_\text{BS3})\cos(\phi_1+\phi)\cos \theta , \\
	C_z	&= -\cos(\phi_2 + \phi + \phi_\text{BS3})\sin \theta .
\end{align}
\end{subequations}
The signal slope is therefore
\begin{align}
	\frac{\partial \langle \hat{J}_z^\text{out} \rangle}{\partial g}	&= \frac{\partial C_x}{\partial g} \langle \hat{J}_x^\text{OAT} \rangle + \frac{\partial C_y}{\partial g} \langle \hat{J}_y^\text{OAT} \rangle + \frac{\partial C_z}{\partial g} \langle \hat{J}_z^\text{OAT} \rangle,
\end{align}
where
\begin{subequations}
\begin{align}
	\frac{\partial C_x}{\partial g} 	&= -k_0 \Big[\left( T^2 - \cos \theta  T_\text{OAT}^2\right)\cos(k_0 g T^2 + \phi + \phi_\text{BS3})\cos(k_0 g T_\text{OAT}^2+\phi) \notag \\
							&+ \left(  \cos \theta  T^2 -T_\text{OAT}^2\right) \sin(k_0 g T^2 + \phi + \phi_\text{BS3})\sin(k_0 g T_\text{OAT}^2+\phi)\Big], \\ 
	\frac{\partial C_y}{\partial g}	&= k_0 \Big[\left( T^2 - \cos \theta  T_\text{OAT}^2\right)\cos(k_0 g T^2 + \phi + \phi_\text{BS3})\sin(k_0 g T_\text{OAT}^2+\phi) \notag \\
							&- \left(  \cos \theta  T^2 -T_\text{OAT}^2\right) \sin(k_0 g T^2 + \phi + \phi_\text{BS3})\cos(k_0 g T_\text{OAT}^2+\phi)\Big], \\ 
	\frac{\partial C_z}{\partial g}	&= k_0 T^2 \sin(k_0 g T^2 + \phi + \phi_\text{BS3})\sin \theta .
\end{align}
\end{subequations}
The variance $\text{Var}(\hat{J}_z^\text{out})$ can also be calculated using Eq.~(\ref{Jz_out_exp}) and Eqs.~(\ref{Jz_out_coeffs}), although we do not present the full expression here as it is not particularly illuminating. However, choosing $\phi = - k_0 g T_\text{OAT}^2 + \phi' $ and $\phi_\text{BS3} = - k_0 g (T^2 - T_\text{OAT}^2) - \phi'$ gives the simplified expressions
\begin{align}
	\frac{\partial \langle \hat{J}_z^\text{out} \rangle}{\partial g}	&= -k_0 ( T^2 -  \cos \theta  T_\text{OAT}^2 ) \langle \hat{J}_\perp^\text{OAT}(\phi' ) \rangle, \\
	\text{Var}(\hat{J}_z^\text{out})	&= \sin^2 \theta  \text{Var}(\hat{J}_z^\text{OAT}) + \cos^2 \theta  \text{Var}(\hat{J}_\parallel^\text{OAT}(\phi' )) - 2 \cos \theta  \sin \theta  \text{Cov}(\hat{J}_\parallel^\text{OAT}(\phi'), \hat{J}_z^\text{OAT}),
\end{align}
where 
\begin{subequations}
\begin{align}
	\hat{J}_\parallel^\text{OAT}(\phi' ) 	&= \sin \phi'  \hat{J}_x^\text{OAT} + \cos \phi'  \hat{J}_y^\text{OAT}, \\
	\hat{J}_\perp^\text{OAT}(\phi' ) 		&= \sin (\phi' +\pi/2) \hat{J}_x^\text{OAT} + \cos(\phi' +\pi/2) \hat{J}_y^\text{OAT},
\end{align}
\end{subequations}
and $\text{Cov}(\hat{X},\hat{Y}) = \langle \hat{X} \hat{Y} + \hat{Y} \hat{X} \rangle /2 - \langle \hat{X} \rangle \langle \hat{Y} \rangle$ is the symmetrized covariance of operators $\hat{X}$ and $\hat{Y}$. Substituting into Eq.~(\ref{sensitivity_formula}) we arrive at the sensitivity
\begin{equation}
	\Delta g = \frac{\xi_{\theta,\phi'}}{\sqrt{N} k_0|T^2 - \cos \theta  T_\text{OAT}^2|} \approx \frac{\xi_{\theta,\phi'}}{\sqrt{N} k_0 T^2},
\end{equation}  
where the above approximation holds in the $T \gg T_\text{OAT}$ regime and
\begin{align}
	\xi_{\theta,\phi'}^2 	&= N \frac{\text{Var}(\hat{J}_{\theta,\phi' }^\text{OAT})}{\langle \hat{J}_{\frac{\pi}{2},\phi' +\frac{\pi}{2}}^\text{OAT}\rangle^2}, \label{defn_xi}
\end{align}
with $\hat{J}_{\theta,\phi'}^\text{OAT} = \sin \theta \sin \phi' \hat{J}_x^\text{OAT} + \sin \theta \cos \phi' \hat{J}_y^\text{OAT} + \cos \theta \hat{J}_z^\text{OAT}$. Optimising over $\theta$ and $\phi'$, achieved by optimally choosing the beamsplitting angle and phase of BS2, gives the minimum sensitivity reported as Eq.~(1) of the main text.

\section{Derivation of analytic model of spin squeezing}
In this section we provide details of the simple analytic model of spin squeezing reported in the main text and used to derive Eq.~(3).

\subsection{Linear ansatz for field operators}
We assume the following ansatz, justified in Ref.~[75]
\begin{subequations}
\label{single_mode_ansatz}
\begin{align}
	\hat{\psi}_1(\textbf{r}) 	&= u_1(\textbf{r}) \hat{a}_1 + \hat{v}_1(\textbf{r}), \\
	\hat{\psi}_2(\textbf{r}) 	&= u_2(\textbf{r}) e^{i k_0 z} \hat{a}_2 + \hat{v}_2(\textbf{r}),
\end{align}
\end{subequations}
where $\int d\textbf{r} \, |u_i(\textbf{r})|^2 = 1$ and $\hat{v}_i(\textbf{r})$ are `vacuum' operators satisfying $\hat{v}_i(\textbf{r}) |\Psi\rangle = 0$. This is akin to assuming that our quantum state is of the form
\begin{equation}
	|\Psi\rangle = \sum_{n_1,n_2 = 0}^\infty C_{n_1,n_2} \frac{(\hat{a}_1^\dag)^{n_1}}{\sqrt{n_1}}\frac{(\hat{a}_2^\dag)^{n_2}}{\sqrt{n_2}}|\text{vac}\rangle,
\end{equation}
for complex coefficients $C_{n_1,n_2}$. This assumption on the state, alongside the field operators' commutation relations, imply that $\hat{v}_i(\textbf{r})$ satisfy
\begin{equation}
	[ \hat{v}_i(\textbf{r}), \hat{v}_j^\dag(\textbf{r}) ] = \delta_{ij} \left( \delta(\textbf{r}-\textbf{r}') - u_i(\textbf{r}) u_j^*(\textbf{r}')\right).
\end{equation}
Substituting ansatz Eqs.~(\ref{single_mode_ansatz}) into the pseudospin operator expressions Eq.~(\ref{pseudo_spin_operators}) and applying $\hat{v}_i(\textbf{r}) |\Psi\rangle = 0$, we obtain the following expectations
\begin{subequations}
\label{Big_J_expressions}
\begin{align}
	\langle \hat{J}_x^\text{OAT} \rangle	&= |\mathcal{Q}| \big(  \cos \varphi \langle \hat{j}_x^\text{OAT} \rangle - \sin \varphi \langle \hat{j}_y^\text{OAT} \rangle \big), \\
	\langle \hat{J}_y^\text{OAT} \rangle	&= |\mathcal{Q}| \big(  \sin \varphi \langle \hat{j}_x^\text{OAT} \rangle + \cos \varphi \langle \hat{j}_y^\text{OAT} \rangle \big), \\
	\langle \hat{J}_z^\text{OAT} \rangle	&= \langle \hat{j}_z^\text{OAT} \rangle, \\
	\langle (\hat{J}_x^\text{OAT})^2 \rangle	&= |\mathcal{Q}|^2 \big\langle ( \cos \varphi \hat{j}_x^\text{OAT} - \sin \varphi \hat{j}_y^\text{OAT} )^2 \big\rangle + \tfrac{1}{4}\left(1 - |\mathcal{Q}|^2\right) \langle \hat{n}^\text{OAT} \rangle, \\
	\langle (\hat{J}_y^\text{OAT})^2 \rangle	&= |\mathcal{Q}|^2 \big\langle ( \sin \varphi \hat{j}_x^\text{OAT} + \cos \varphi \hat{j}_y^\text{OAT} )^2\big\rangle + \tfrac{1}{4}\left(1 - |\mathcal{Q}|^2\right) \langle \hat{n}^\text{OAT} \rangle, \\
	\langle (\hat{J}_z^\text{OAT})^2 \rangle	&= \langle (\hat{j}_z^\text{OAT})^2 \rangle, \\
	\tfrac{1}{2}\langle \hat{J}_x^\text{OAT} \hat{J}_y^\text{OAT} + \hat{J}_y^\text{OAT} \hat{J}_x^\text{OAT} \rangle	&= \tfrac{1}{2}|\mathcal{Q}|^2 \big\langle ( \cos \varphi \hat{j}_x^\text{OAT} - \sin \varphi \hat{j}_y^\text{OAT} )( \sin \varphi \hat{j}_x^\text{OAT} + \cos \varphi \hat{j}_y^\text{OAT} ) + h.c. \big\rangle, \\
	\tfrac{1}{2}\langle \hat{J}_x^\text{OAT} \hat{J}_z^\text{OAT} + \hat{J}_z^\text{OAT} \hat{J}_x^\text{OAT} \rangle	&= \tfrac{1}{2}|\mathcal{Q}| \big\langle ( \cos \varphi \hat{j}_x^\text{OAT} - \sin \varphi \hat{j}_y^\text{OAT} ) \hat{j}_z^\text{OAT} + h.c. \big\rangle, \\
	\tfrac{1}{2}\langle \hat{J}_y^\text{OAT} \hat{J}_z^\text{OAT} + \hat{J}_z^\text{OAT} \hat{J}_y^\text{OAT} \rangle	&= \tfrac{1}{2}|\mathcal{Q}| \big\langle ( \sin \varphi \hat{j}_x^\text{OAT} + \cos \varphi \hat{j}_y^\text{OAT} ) \hat{j}_z^\text{OAT} + h.c. \big\rangle,
\end{align}
\end{subequations}
where
\begin{subequations}
\begin{align}
	\hat{j}_x	&= \tfrac{1}{2}\left( \hat{a}_1^\dag \hat{a}_2 + \hat{a}_1 \hat{a}_2^\dag \right), \\
	\hat{j}_y	&= -\tfrac{i}{2}\left( \hat{a}_1^\dag \hat{a}_2 - \hat{a}_1 \hat{a}_2^\dag \right), \\
	\hat{j}_z	&= \tfrac{1}{2}\left( \hat{a}_1^\dag \hat{a}_1 - \hat{a}_2^\dag \hat{a}_2 \right), \\
	\hat{n}	&=\hat{a}_1^\dag \hat{a}_1 + \hat{a}_2^\dag \hat{a}_2,
\end{align}
\end{subequations}
$\hat{j}_i^\text{OAT} = \hat{U}_\text{nonlinear}^\dag \hat{j}_i \hat{U}_\text{nonlinear}$, $\hat{n}^\text{OAT} = \hat{U}_\text{nonlinear}^\dag \hat{n} \hat{U}_\text{nonlinear}$, and the complex spatial-mode overlap $\mathcal{Q} = |\mathcal{Q}| e^{i \varphi}$ is
\begin{equation}
	\mathcal{Q} = \int d\textbf{r} \, u_1^*(\textbf{r}) u_2(\textbf{r}). \label{complex_mode_overlap}
\end{equation}
This allows us to write Eq.~(\ref{defn_xi}) as
\begin{align}
	\xi_{\theta,\phi}^2	&= N \frac{\sin^2 \theta \text{Var}(\hat{J}_z^\text{OAT}) + \cos^2 \theta \text{Var}(\hat{J}_{\frac{\pi}{2},\phi}^\text{OAT}) - 2 \cos \theta \sin \theta \text{Cov}(\hat{J}_{\frac{\pi}{2},\phi}^\text{OAT},\hat{J}_z^\text{OAT})}{\langle \hat{J}_{\frac{\pi}{2},\phi+\frac{\pi}{2}}^\text{OAT} \rangle^2}, \\
					&= N \frac{\sin^2 \theta \text{Var}(\hat{j}_z^\text{OAT}) + \cos^2 \theta \left( |\mathcal{Q}|^2 \text{Var}(\hat{j}_{\frac{\pi}{2},\phi + \varphi}^\text{OAT}) + \frac{1}{4}\left( 1 - |\mathcal{Q}|^2\right)\langle \hat{n}^\text{OAT} \rangle \right) - \sin (2\theta) |\mathcal{Q}| \text{Cov}(\hat{j}_{\frac{\pi}{2},\phi + \varphi}^\text{OAT},\hat{j}_z^\text{OAT})}{|\mathcal{Q}|^2\langle \hat{j}_{\frac{\pi}{2},\phi + \varphi+\frac{\pi}{2}}^\text{OAT} \rangle^2}, \label{general_spin_squeezing_parameter}
\end{align}
where $\hat{j}_{\theta,\phi}^\text{OAT} = \sin \theta \sin \phi \hat{j}_x^\text{OAT} + \sin \theta \cos \phi \hat{j}_y^\text{OAT} + \cos \theta \hat{j}_z^\text{OAT}$. A non-zero phase $\varphi$ induces a drift of the overall pseudospin vector along the equator of the Bloch sphere, which reduces the average pseudospin length, and consequently the degree of squeezing. Explicitly, 
\begin{align}
	\langle \hat{J}_{\frac{\pi}{2},\phi + \frac{\pi}{2}}^\text{OAT} \rangle 	&= \cos \phi \langle \hat{J}_x^\text{OAT} \rangle - \sin \phi \langle \hat{J}_y^\text{OAT} \rangle = |\mathcal{Q}| \left(\cos \left( \phi + \varphi \right) \langle \hat{j}_x^\text{OAT} \rangle - \sin \left( \phi + \varphi \right) \langle \hat{j}_y^\text{OAT} \rangle \right) \leq \tfrac{N}{2}. \label{pseudospin_length}
\end{align}
We can compensate for this via the choice $\phi = -\varphi$, which centers the state along the $J_x$-axis. This optimal choice of $\phi$ gives spin squeezing parameter
\begin{align}
	\xi_{\theta,-\varphi}^2 &= N \frac{\sin^2 \theta \text{Var}(\hat{j}_z^\text{OAT}) + \cos^2 \theta \left( |\mathcal{Q}|^2 \text{Var}(\hat{j}_y^\text{OAT}) + \frac{1}{4}\left( 1 - |\mathcal{Q}|^2\right)\langle \hat{n}^\text{OAT} \rangle \right) - \sin (2\theta) |\mathcal{Q}| \text{Cov}(\hat{j}_y^\text{OAT},\hat{j}_z^\text{OAT})}{|\mathcal{Q}|^2\langle \hat{j}_x^\text{OAT} \rangle^2}. \label{xi_optphi}
\end{align}

\subsection{One-axis twisting due to interatomic interactions} {\label{OAT_subsection}}
In order to compute the expectation values in Eq.~(\ref{xi_optphi}), we now derive a simplified model for the BEC evolution between pulses, $\hat{U}_\text{evo}(t,t_0)$. Substituting ansatz Eqs.~(\ref{single_mode_ansatz}) into Hamiltonian Eq.~(\ref{full_field_Hamiltonian}) and neglecting operators $\hat{v}_i(\textbf{r})$ and the kinetic energy term:
\begin{align}
	\hat{H}	&\approx \hat{H}(t) = \hbar \chi_{11}(t) \hat{n}_1(\hat{n}_1 - 1) + \hbar \chi_{22}(t) \hat{n}_2(\hat{n}_2-1) + 2 \hbar \chi_{12}(t) \hat{n}_1\hat{n}_2, \notag \\
			&= \hbar \chi(t) \hat{j}_z^2 + \hbar (\chi_{11}(t)-\chi_{22}(t))(\hat{n} - 1) \hat{j}_z  - \tfrac{1}{2}\hbar (\chi_{11}(t) + \chi_{22}(t))\hat{n}+ \tfrac{1}{4}\hbar(\chi_{11}(t)+\chi_{22}(t) + 2\chi_{12}(t))\hat{n}^2, \label{single_mode_Ham}
\end{align}
where $\hat{n}_i = \hat{a}_i^\dag \hat{a}_i$, $\hat{n} = \hat{n}_1 + \hat{n}_2$, and $\chi(t) = \chi_{11}(t) + \chi_{22}(t) - 2\chi_{12}(t)$ with $\chi_{ij}(t) = \frac{g_{ij}}{2 \hbar} \int d\textbf{r} \, |u_i(\textbf{r},t)|^2 |u_j(\textbf{r},t)|^2$. Although we have neglected the kinetic energy term from Hamiltonian~Eq.~(\ref{full_field_Hamiltonian}), we still account for the effect of the kinetic energy via the time-dependence of $u_i(\textbf{r},t)$, which we impose \emph{a posteriori} (e.g. by identifying $|u_i(\textbf{r},t)|^2$ as the normalized condensate density of state $|i\rangle$, determined via solution of the GPE). This does neglect a relative phase shift accrued between the two modes during free propagation. However, the symmetry of the interferometer ensures that this overall phase shift is zero at the output.

For our analytic model we work within the SU(2) algebra where the total atom number $\hat{n}$ is a constant of motion. The third and fourth terms in Eq.~(\ref{single_mode_Ham}) therefore have no effect on the evolution and can be neglected. The second term in Eq.~(\ref{single_mode_Ham}) results in a rotation about $\hat{j}_z$. However, since $\chi_{11}(t) \approx \chi_{22}(t)$ for the $|F=1,m_F=0\rangle$ and $|F=2,m_F=0\rangle$ hyperfine states of $^{87}$Rb, this rotation will be negligible. More generally, the rotation could be compensated for by including an additional phase shift on BS2. Therefore, the Hamiltonian Eq.~(\ref{single_mode_Ham}) approximately corresponds to the OAT Hamiltonian $\hat{H}_\text{OAT}(t) = \hbar \chi(t) \hat{j}_z^2$, giving
\begin{equation}
	\hat{U}_\text{evo}(t,t_0) \approx \hat{U}_\text{OAT}(t,t_0) =  \exp\left[ -\frac{i}{\hbar} \int_{t_0}^t dt' \hat{H}_\text{OAT}(t')\right] = \exp\left[ -i \lambda(t,t_0) \hat{j}_z^2 \right],
\end{equation}
where $ \lambda(t,t_0) \equiv \int_{t_0}^t dt' \chi(t')$. 

Our scheme uses two periods of OAT evolution, separated by a $\pi$-pulse (operations 2-4 above). This formally corresponds to a single period of OAT evolution $\hat{U}_\text{OAT}(2T_\text{OAT},0) = \exp[-i \lambda(2T_\text{OAT},0)\hat{j}_z^2]$. To see this, first note that $\hat{U}_{\pi,0}^\dag \hat{j}_z \hat{U}_{\pi,0} = -\hat{j}_z$. Then
\begin{align}
	\hat{U}_\text{OAT}(2T_\text{OAT},T_\text{OAT}) \hat{U}_{\pi,0} \hat{U}_\text{OAT}(T_\text{OAT},0)	&=  \hat{U}_{\pi,0}  \hat{U}_{\pi,0}^\dag e^{-i \lambda(2T_\text{OAT},T_\text{OAT}) \hat{j}_z^2} \hat{U}_{\pi,0} e^{-i \lambda(T_\text{OAT},0) \hat{j}_z^2} \notag \\
																				&= \hat{U}_{\pi,0} e^{-i \lambda(2T_\text{OAT},T_\text{OAT}) [\hat{U}_{\pi,0}^\dag\hat{j}_z \hat{U}_{\pi,0}]^2} e^{-i \lambda(T_\text{OAT},0) \hat{j}_z^2} \notag \\
																				&= \hat{U}_{\pi,0} e^{-i [\lambda(2T_\text{OAT},T_\text{OAT}) + \lambda(T_\text{OAT},0)] \hat{j}_z^2}.
\end{align}
The unitary $\hat{U}_{\pi,0}$ can be neglected since it has no meaningful effect on the interferometer sequence; it is equivalent to an additional $\pi$ phase shift on BS2 which can be compensated for by redefining $\theta \to \theta + \pi$. The remaining unitary clearly corresponds to a single period of OAT evolution from $t=0$ to $t=2T_\text{OAT}$, since $\lambda(2T_\text{OAT},T_\text{OAT}) + \lambda(T_\text{OAT},0) = \lambda(2T_\text{OAT},0)$.

\subsection{Derivation of minimum spin squeezing parameter, Eq.~(3) of main text}
We have shown that $\hat{U}_\text{nonlinear} \approx \hat{U}_\text{OAT}(2T_\text{OAT},0) \hat{U}_{\pi/2,-\pi/2}$ and are now in a position to calculate the expectations in Eq.~(\ref{xi_optphi}). Denoting $\lambda \equiv \lambda(2T_\text{OAT},0)$, OAT evolves the pseudospin operators as~[59]
\begin{subequations}
\label{OAT_transformation_exact}
\begin{align}
	\hat{j}_x^\text{OAT}(\lambda) 	&= \frac{1}{2} \left( \hat{j}_+^{(1)} e^{i 2 \lambda (\hat{j}_z^{(1)} + \frac{1}{2})} + e^{-i 2 \lambda (\hat{j}_z^{(1)} + \frac{1}{2})} \hat{j}_-^{(1)}\right), \\
	\hat{j}_y^\text{OAT}(\lambda) 	&= -\frac{i}{2} \left( \hat{j}_+^{(1)} e^{i 2 \lambda (\hat{j}_z^{(1)} + \frac{1}{2})} - e^{-i 2 \lambda (\hat{j}_z^{(1)} + \frac{1}{2})} \hat{j}_-^{(1)}\right), \\
	\hat{j}_z^\text{OAT}(\lambda)	&= \hat{j}_z^{(1)} ,
\end{align}
\end{subequations}
where $\hat{j}_\pm^{(1)} = \hat{j}_x^{(1)} \pm i \hat{j}_y^{(1)}$ and the superscript `(1)' signifies that expectations of these operators are taken with respect to the state immediately after the first beamsplitter. i.e. $\hat{j}_i^{(1)} = \hat{U}_{\pi/2,-\pi/2}^\dag \hat{j}_i \hat{U}_{\pi/2,-\pi/2}$. Since this state is a $\hat{j}_x$ eigenstate, we obtain the following expectations~[59]
\begin{subequations}
\label{OAT_squeezing_expectations}
\begin{align}
	\langle \hat{n} \rangle	&= N, \\
	\langle \hat{j}_x^\text{OAT}\rangle	&= \frac{N}{2} \cos^{N-1} \lambda, \\ 
	\langle \hat{j}_y^\text{OAT}\rangle	&= 0, \\
	\langle \hat{j}_z^\text{OAT}\rangle	&= 0, \\
	\langle (\hat{j}_x^\text{OAT})^2\rangle	&= \frac{N}{8}\left[ N + 1 + (N-1) \cos^{N-2}(2\lambda) \right], \notag \\
									&\approx \frac{N^2}{8}\left[ 1 + \cos^N(2\lambda) \right], \\
	\langle (\hat{j}_y^\text{OAT})^2\rangle	&= \frac{N}{4}\left[ 1+\tfrac{1}{2}(N-1)(1-\cos^{N-2}(2\lambda))\right], \notag \\
									&\approx \frac{N}{4}\left[ 1+\tfrac{1}{2}N(1-\cos^N(2\lambda))\right], \\
	\langle (\hat{j}_z^\text{OAT})^2\rangle	&= \frac{N}{4}, \\
	\tfrac{1}{2}\langle \hat{j}_x^\text{OAT} \hat{j}_y^\text{OAT} + \hat{j}_y^\text{OAT} \hat{j}_x^\text{OAT} \rangle 	& = 0, \\
	\tfrac{1}{2}\langle \hat{j}_x^\text{OAT} \hat{j}_z^\text{OAT} + \hat{j}_z^\text{OAT} \hat{j}_x^\text{OAT} \rangle 	&= \frac{1}{2}\frac{\partial}{\partial \lambda}\langle \hat{j}_y^\text{OAT} \rangle = 0, \\
	\tfrac{1}{2}\langle \hat{j}_y^\text{OAT} \hat{j}_z^\text{OAT} + \hat{j}_z^\text{OAT} \hat{j}_y^\text{OAT} \rangle 	&= -\frac{1}{2}\frac{\partial}{\partial \lambda}\langle \hat{j}_x^\text{OAT} \rangle = \frac{N}{4}(N-1) \sin \lambda \cos^{N-2} \lambda, \notag \\
									&\approx \frac{N^2}{4} \sin \lambda \cos^N \lambda.
\end{align}
\end{subequations}
Substituting these expressions into Eq.~(\ref{xi_optphi}) gives
\begin{equation}
	\xi_{\theta,-\varphi}^2	= N \frac{2 \sin^2 \theta + \left[ 2 + |\mathcal{Q}|^2 N \left( 1 - \cos^N(2 \lambda)\right)\right]\cos^2 \theta -4 \mathcal{Q} N \sin \lambda \cos^N \lambda \cos \theta \sin \theta}{2 |\mathcal{Q}|^2 N \cos^{2N} \lambda}.
\end{equation}
Solving for the optimal $\theta$ gives
\begin{subequations}
\begin{align}
	\theta_\text{sq}		&= \frac{3\pi}{2} + \frac{1}{2} \tan^{-1} \left[ \frac{4 \sin \lambda \cos^N \lambda}{|\mathcal{Q}|\left( \cos^N(2 \lambda)-1\right)}\right], \label{theta_opt_sq}\\
	\theta_\text{a-sq}	&= \pi+ \frac{1}{2} \tan^{-1} \left[ \frac{4 \sin \lambda \cos^N \lambda}{|\mathcal{Q}|\left( \cos^N(2 \lambda)-1\right)}\right],
\end{align}
\end{subequations}
which are the angles that give minimum squeezing and maximum anti-squeezing, respectively. Explicitly, the minimum squeezing is
\begin{equation}
	\xi_{\theta_\text{sq},-\varphi}^2 = N \frac{4 + |\mathcal{Q}|^2 N \left( 1 - \cos^N(2 \lambda) \right)\left(1 - \sqrt{1 + \frac{16 \sin^2 \lambda \cos^{2N} \lambda}{|\mathcal{Q}|^2\left( 1 - \cos^N(2\lambda) \right)^2}} \right)}{4|\mathcal{Q}|^2 N \cos^{2N} \lambda}.
\end{equation}
In the linear squeezing regime we can approximate
\begin{subequations}
\label{eqs_linear_squeezing}
\begin{align}
	\left( 1 - \cos^N(2 \lambda) \right)	&\approx 2 N \lambda^2, \\
	\frac{16 \sin^2 \lambda \cos^{2N} \lambda}{|\mathcal{Q}|^2\left( 1 - \cos^N(2\lambda) \right)}	&\approx \frac{4}{|\mathcal{Q}|^2 N^2 \lambda^2}, \\
	\cos^{2N}\lambda 	&\approx 1,
\end{align}
\end{subequations}
yielding $\theta_\text{sq} \approx \tfrac{3\pi}{2} - \frac{1}{2} \tan^{-1} \left[2/(N |\mathcal{Q}| \lambda) \right]$ and Eq.~(3) of the main text.

\section{Gross-Pitaevskii Equation Numerical Simulations} \label{GPE_sec}
Data for $\mathcal{Q}$, $\lambda$, and the mean-field densities plotted in Figs 2 and 3 of the main text were generated by numerically simulating the two component Gross-Pitaevskii equation (GPE)
\begin{subequations} \label{2compGPE}
\begin{align}
	i \hbar \frac{\partial}{\partial t}\Psi_1(\textbf{r},t) = \left[H(\textbf{r}) + g_{11} |\Psi_1(\textbf{r},t)|^2 + g_{12} |\Psi_2(\textbf{r},t)|^2\right] \Psi_1(\textbf{r},t), \\
	i \hbar \frac{\partial}{\partial t}\Psi_2(\textbf{r},t) = \left[H(\textbf{r}) + g_{12} |\Psi_1(\textbf{r},t)|^2 + g_{22} |\Psi_2(\textbf{r},t)|^2\right] \Psi_2(\textbf{r},t),
\end{align}
\end{subequations}
where $H(\textbf{r}) = -\tfrac{\hbar^2}{2m}\nabla^2 + m g z$ and $\Psi_i(\textbf{r},t)$ is the mean-field condensate wavefunction for atoms in state $|i\rangle$. We simulated Eqs~(\ref{2compGPE}) using the open-source software package XMDS2~[78] 
with an adaptive 4th-5th order Runge-Kutta interaction picture algorithm under the assumption of cylindrical symmetry (i.e. $\Psi_i(\textbf{r}) = \Psi_i(r_\perp,z)$ where $r_\perp^2 = x^2 + y^2$), thereby allowing the efficient computation of derivatives via Hankel transforms. Imaginary time propagation~[79] 
was used to find the GPE groundstate for a given atom number $N$ and trapping potential $V(r_\perp,z) = \frac{1}{2} m \omega_\perp r_\perp^2 + \frac{1}{2} m \omega_z z^2$, with $(\omega_\perp,\omega_z) = 2\pi \times (50,50)$ Hz and $(\omega_\perp,\omega_z) = 2\pi \times (32,160)$ Hz for the spherical and pancake BEC cases, respectively. Since component 1 and 2 are ideally centered around $(k_{r_\perp}, k_z) = (0,0)$ and $(k_{r_\perp}, k_z) = (0, 2k_0)$, respectively, in $k$-space, a further computational efficiency was obtained by making the transformation $\tilde \Psi_2(\textbf{r},t) = \exp(-i k_0 z) \Psi_2(\textbf{r},t)$ in Eqs.~(\ref{2compGPE}). This centers both components $(k_{r_\perp}, k_z) = (0, 0)$, enabling simulations with much smaller $k$-space grids (and therefore much fewer grid points). Explicitly, our simulations required grid points $(N_{r_\perp},N_z)$ between $(64, 256)$ and $(160,1024)$. Simulations were conducted in the freely-falling frame where $g = 0$ and beamsplitters were treated as instantaneous linear transformations, as described in the main text.

\section{Truncated Wigner Stochastic Numerical Simulations}
The derivation of the truncated Wigner (TW) method has been described in detail elsewhere~[76,80--82]. 
Briefly, the system's evolution can be written as a partial differential equation (PDE) for the system's Wigner function by exploiting correspondences between differential operators on the Wigner function and the original quantum operators~[83, 84]. 
Once third- and higher-order derivatives are truncated (an approximation that is typically valid provided the occupation per mode is not too small for appreciable time periods~[77]), 
this PDE takes the form of a Fokker-Planck equation, which can be efficiently simulated by a set of stochastic differential equations (SDEs). For our case, the TW SDEs that simulate evolution under full-field Hamiltonian Eq.~(\ref{full_field_Hamiltonian}) are
\begin{subequations} \label{2compTW}
\begin{align}
	i \hbar \frac{\partial}{\partial t}\Phi_1(\textbf{r},t) = \left[H(\textbf{r}) + g_{11} \left(|\Phi_1(\textbf{r},t)|^2 - \frac{1}{\Delta V} \right) + g_{12} \left(|\Phi_2(\textbf{r},t)|^2 - \frac{1}{2\Delta V}\right)\right] \Phi_1(\textbf{r},t), \\
	i \hbar \frac{\partial}{\partial t}\Phi_2(\textbf{r},t) = \left[H(\textbf{r}) + g_{12} \left(|\Phi_1(\textbf{r},t)|^2 - \frac{1}{2\Delta V} \right) + g_{22} \left(|\Phi_2(\textbf{r},t)|^2  - \frac{1}{\Delta V} \right)\right] \Phi_2(\textbf{r},t),
\end{align}
\end{subequations}
where $H(\textbf{r}) = -\tfrac{\hbar^2}{2m}\nabla^2 + m g z$ and $\Delta V$ is the volume element of the simulation spatial grid. The complex fields $\Phi_i(\textbf{r},t)$ loosely correspond to the field operators $\hat{\psi}_i(\textbf{r},t)$; formally, expectation values of some arbitrary operator function $f$ are computed by averaging over solutions to Eqs.~(\ref{2compTW}) with the stochastically sampled initial conditions. Explicitly $\langle \{f[\hat{\psi}_1,\hat{\psi}_2]\}_\text{sym} \rangle = \overline{f(\Phi_1,\Phi_2)}$, where `sym' denotes symmetric ordering. For example, $\langle \hat{\psi}_1^\dag(\textbf{r})\hat{\psi}_2(\textbf{r}')\rangle = \overline{\Phi_1^*(\textbf{r})\Phi_2(\textbf{r}')}$, $\langle \hat{\psi}_i^\dag(\textbf{r})\hat{\psi}_i(\textbf{r})\rangle = \overline{|\Phi_i(\textbf{r})|^2} - \tfrac{1}{2\Delta V}$, and $\langle \hat{\psi}_i^\dag(\textbf{r})\hat{\psi}_i^\dag(\textbf{r})\hat{\psi}_i(\textbf{r})\hat{\psi}_i(\textbf{r})\rangle = \overline{|\Phi_i(\textbf{r})|^4} - \tfrac{2}{\Delta V}\overline{|\Phi_i(\textbf{r})|^2} + \tfrac{1}{2 \Delta V^2}$.

The initial conditions for the SDEs Eqs.~(\ref{2compTW}) are randomly sampled from the Wigner distribution of the initial quantum state. Initially, all atoms in the BEC are in internal state $|1\rangle$, which we model as a multimode coherent state $|\Psi_1(0)\rangle = 	\exp[\sqrt{N}(\hat{a}_{\Psi_1} - \hat{a}_{\Psi_1}^\dag)] |\text{vac}\rangle$ where $\hat{a}_{\Psi_1} = \frac{1}{\sqrt{N}}\int d\textbf{r}\, \Psi_1^*(\textbf{r},0) \hat{\psi}_1(\textbf{r})$ 
and $\Psi_1(\textbf{r},0)$ is the GPE groundstate under harmonic confinement, obtained via imaginary time evolution. Internal state $|2\rangle$ is entirely unoccupied and therefore $|\Psi_2(0)\rangle$ is a vacuum state. This initial condition $|\Psi(0)\rangle = | \Psi_1(0)\rangle \otimes |\Psi_2(0)\rangle$ is sampled via $\Phi_1(\textbf{r},0) =  \Psi_1(\textbf{r},0) + \eta_1(\textbf{r})$ and $\Phi_2(\textbf{r},0) = \eta_2(\textbf{r})$, where $\eta_i(\textbf{r})$ are complex Gaussian noises with mean zero and $\overline{\eta_i^*(\textbf{r}_n)\eta_j(\textbf{r}_m)} = \tfrac{1}{2 \Delta V} \delta_{ij} \delta_{nm}$ for spatial grid points $\textbf{r}_m$ and $\textbf{r}_n$~[85]. 
This initial condition neglects the effect of thermal fluctuations, which is an excellent approximation for typical ultracold-atom gravimeters, as discussed in the section `Justification for zero-temperature model' below.

The data shown in Fig.~3 of the main text was generated from simulations of Eqs.~(\ref{2compTW}) using a simulation procedure and parameters similar to that described for Eqs.~(\ref{2compGPE}). Acceptable sampling errors required the simulation of between 2,000 and 30,000 stochastic trajectories.

\section{Effective 1D TW Stochastic Numerical Simulations}
Data presented in Figs 3,4, and 5 of the main text were obtained by an effective 1D TW simulation that captures the free expansion dynamics in the radial co-ordinate via a time-dependent 1D interaction strength. Here we provide a brief derivation of this model and provide additional details on the simulation procedure for the full interferometer sequence used to directly compute $\Delta g$.

\subsection{Derivation of effective 1D model}
For simplicity, we present this derivation for a single-component GPE, however it trivially generalizes to multiple components and the SDEs for the TW method. 

Assume that a single-component BEC is initially prepared in a cylindrically-symmetric harmonic potential with axial and radial trapping frequencies $\omega_z$ and $\omega_\perp$, respectively. If the trap is turned off then the mean-field dynamics of the condensate are governed by the GPE
\begin{equation}
	i \hbar \frac{\partial \Psi}{\partial t} = \left( -\frac{\hbar^2}{2m} \nabla^2 \Psi + V(z,t) + g_{3\text{D}} |\Psi|^2 \right) \Psi, \label{3D_GPE}
\end{equation}
where potential $V(z,t)$ predominantly modifies the centre-of-mass motion of the BEC in the $z$ direction (i.e. it does not strongly affect the expansion dynamics in the $z$ direction). This is the operating regime of our atom interferometer. In this regime, the BEC's radial profile will undergo self-similar expansion when released from the trap. This motivates the ansatz
\begin{equation}
	\Psi(z,r_\perp,t) = \psi(z,t) \Phi_\text{TF}(r_\perp, t), \label{scaling_ansatz}
\end{equation}
where $\Phi_\text{TF}(r_\perp, t) = \sqrt{\rho_\text{TF}(r_\perp,t)} e^ {i S(r_\perp,t)} e^{-i\beta(t)}$ is the radial Thomas-Fermi (TF) solution to the free expansion dynamics, normalized to unity:
\begin{subequations}
\begin{align}
	\rho_\text{TF}(r_\perp,t)	&= \max \left\{\frac{2}{\pi R_\perp(t)^2}\left[ 1 - \left(\frac{r_\perp}{R_\perp(t)}\right)^2 \right], 0 \right\}, \\
	S(r_\perp,t)	&= \frac{m}{2 \hbar} \frac{\dot{b}_\perp(t)}{b_\perp(t)} r_\perp^2.
\end{align}
\end{subequations}
Here $\beta(t)$ is a global phase factor we can choose arbitrarily and $b_\perp(t)$ scales the initial TF radius of the cloud according to $R_\perp(t) = b_\perp(t) R_\perp(0)$, where the initial width $R_\perp(0) = \sqrt{2 \mu / (m \omega_\perp^2)}$ is determined by the chemical potential $\mu$ of the initial TF groundstate. We determine $b_\perp(t)$ from the scaling solutions for a freely-expanding BEC released from a cylindrically-symmetric harmonic potential~[69]:
\begin{subequations}
\label{scaling_solns}
\begin{align}
	\ddot{b}_\perp 	&= \frac{\omega_\perp^2}{b_\perp^3 b_z}, \\
	\ddot{b}_z 	&= \frac{\omega_z^2}{b_\perp^2 b_z^2}.
\end{align}
\end{subequations}

Substituting our ansatz Eq.~(\ref{scaling_ansatz}) into Eq.~(\ref{3D_GPE}), multiplying by $\Phi(r_\perp, t)$ and integrating over $r_\perp$ yields
\begin{align}
	i \hbar \frac{\partial \psi}{\partial t} 	&= \left[ -\frac{\hbar^2}{2 m} \frac{\partial^2}{\partial z^2} + V(z,t) + \frac{4 g_{3\text{D}}}{3 \pi R_\perp(t)^2}|\psi(z,t)|^2\right] \psi(z,t) - i \hbar \left[ 2\pi \int dr_\perp r_\perp \Phi^*(r_\perp,t) \frac{\partial \Phi (r_\perp,t)}{\partial t} \right] \psi(z,t) \notag \\
								& -\frac{\hbar^2}{2m}\left[ 2\pi \int dr_\perp \Phi^*(r_\perp,t) \frac{\partial}{\partial r_\perp} \left( r_\perp \frac{\partial \Phi(r_\perp,t)}{\partial r_\perp} \right)\right] \psi(z,t). \label{1D_GPE_terms}
\end{align}
The integral in the second term on the right-hand side evaluates to
\begin{equation}
	i \hbar \int dr_\perp r_\perp \Phi^*(r_\perp,t) \frac{\partial \Phi (r_\perp,t)}{\partial t} = \hbar \dot{\beta}(t) + \frac{1}{6} m R_\perp(0)^2 \left( \dot{b}_\perp(t)^2 - b_\perp(t) \ddot{b}_\perp(t) \right).
\end{equation}
It can consequently be set to zero by a judicious choice of $\beta(t)$. The third term on the RHS of Eq.~(\ref{1D_GPE_terms}) is negligible during the initial expansion dynamics, since the Thomas-Fermi profile has negligible kinetic energy. In the latter stage of expansion when most of the interaction energy has been converted to kinetic energy, the radial and axial co-ordinates decouple, and so this term continues to have negligible effect on the dynamics of $\psi(z,t)$, and can be safely neglected. We therefore arrive at an effective 1D GPE with time-dependent interaction strength $g_{1\text{D}}(t) = 4 g_{3\text{D}}/ [3 \pi R_\perp(0)^2 b_\perp(t)^2]$, which is determined by solving Eqs~(\ref{scaling_solns}) in parallel with the effective 1D equation for $\psi(z,t)$.

\subsection{Simulation procedure for full interferometer sequence}
The effective 1D TW method SDEs are
\begin{subequations} \label{1DTWeom}
\begin{align}
	i \hbar \frac{\partial}{\partial t}\Phi_1^{1\text{D}}(z,t) = \left[H(\textbf{r}) + g_{11}^{1\text{D}}(t) \left(|\Phi_1^{1\text{D}}(z,t)|^2 - \frac{1}{\Delta z} \right) + g_{12}^{1\text{D}}(t)\left(|\Phi_2^{1\text{D}}(z,t)|^2 - \frac{1}{2\Delta z}\right)\right] \Phi_1^{1\text{D}}(z,t), \\
	i \hbar \frac{\partial}{\partial t}\Phi_2^{1\text{D}}(z,t) = \left[H(\textbf{r}) + g_{12}^{1\text{D}}(t) \left(|\Phi_1^{1\text{D}}(z,t))|^2 - \frac{1}{2\Delta z} \right) + g_{22}^{1\text{D}}(t) \left(|\Phi_2^{1\text{D}}(z,t)|^2  - \frac{1}{\Delta z} \right)\right] \Phi_2^{1\text{D}}(z,t),
\end{align}
\end{subequations}
where $\Delta z$ is the $z$-grid spacing of the simulation and $g_{ij}^{1\text{D}}(t) = 4 g_{ij}/ [3 \pi R_\perp(0)^2 b_\perp(t)^2]$ with $R_\perp(0)$ defined above and $b_\perp$ determined by solving Eqs~(\ref{scaling_solns}) in parallel. The stochastic initial conditions are $\Phi_1^{1\text{D}}(z,0) =  \Psi_1^{1\text{D}}(z,0) + \eta_1(z)$ and $\Phi_2^{1\text{D}}(z,0) = \eta_2(z)$, where $|\Psi_1^{1\text{D}}(z,0)|^2 = 2\pi \int_0^\infty dr_\perp \, r_\perp |\Psi_1(r_\perp,z,0)|^2$ and $\eta_i(z)$ are complex Gaussian noises with mean zero and $\overline{\eta_i^*(z_n)\eta_j(z_m)} = \tfrac{1}{2 \Delta z} \delta_{ij} \delta_{nm}$ for spatial grid points $z_m$ and $z_n$. The complex fields $\Phi_i^{1\text{D}}(z,t)$ allow the calculation of expectations such as $2 \pi \int_0^\infty dr_\perp \langle \hat{\psi}^\dag_i(\textbf{r},t)\hat{\psi}_i(\textbf{r},t)\rangle = \overline{|\Phi_i(z,t)|^2} - \tfrac{1}{2\Delta z}$ (i.e. the integrated density of component $i$), and clearly allow the computation of means, variances and covariances of the pseudospin operators.

Simulation of the full interferometer sequence required large domains in $z$ and large grid sizes ($N_z$ between 1024 and 4096 grid points), with very small sampling errors achieved with averages over 10,000 trajectories. The sensitivity was computed via Eq.~(\ref{sensitivity_formula}) with the signal slope approximated as a linear finite difference: $\partial \langle \hat{J}_z^\text{out} \rangle / d g \rangle|_{g=g_0} \approx (\langle \hat{J}_z^\text{out} \rangle|_{g=g_0} - \langle \hat{J}_z^\text{out} \rangle|_{g=g_0 + \delta g}) / \delta g$. This requires two identical TW simulations with $g = g_0$ and $g = g_0 + \delta g$. Without loss of generality, we chose $g_0 = 0$ for all numerical calculations, which is computationally a more efficient choice than larger values of $g$; physically, a large offset in $g$ is easily accounted for by adjusting the beamsplitter phases, as done in typical cold-atom gravimeters~[5]. 
Choosing $\delta g$ anywhere between $10^{-4}$ m/s$^{2}$ and $10^{-10}$ m/s$^{2}$ resulted in approximately the same value for the sensitivity.

\section{Justification for zero-temperature model}
Ultracold-atom gravimeters such as that reported in Ref.~[7] 
use almost pure condensates with no discernible thermal component. In these experiments, the condensate fraction is most likely $> 95\%$, and certainly $> 90\%$. Consequently, the zero temperature initial states used in our analysis provide an excellent description of such ultracold-atom gravimeters.

We can quantitatively confirm that finite temperature effects have minimal impact on our scheme in this regime by employing an initial state sampling procedure similar to that used in Ref.~[86]. 
We use the simple growth stochastic Gross-Pitaevskii equation (SGPE) to sample the grand canonical ensemble of an interacting Bose gas at a given chemical potential and temperature (or equivalently, a given atom number $N$ and condensate fraction $N_c/N$)~[80]. 
This, plus vacuum noise due to quantum fluctuations, provide the initial TW samples for the atoms in $|1\rangle$; the initial state for component $|2\rangle$ is treated as a vacuum state, as in our zero temperature simulations.

Figure~\ref{fig_finite_temp} shows the change in the minimum spin squeezing parameter, $\xi$, for our scheme as a function of condensate fraction for spin squeezing duration $T_\text{OAT} = 10$ms and a total atom number of $N = 10^4$ in a `pancake' geometry (trapping frequencies $f_z = 160$Hz and $f_r = 32$Hz). The condensate fraction of the initial state is determined by computing and diagonalizing the one-body density matrix $\langle \hat{\psi}_1^\dag(z) \hat{\psi}_1(z')\rangle$, with the largest eigenvalue corresponding to the condensate number $N_c$. This shows that finite temperature effects do slightly degrade the spin squeezing (and therefore the sensitivity), with hotter initial states giving a larger degradation (as expected). However, for typical experiments with condensate fractions $>90\%$, our calculation shows that thermal effects only degrade the spin squeezing by at most $10-15$\%.

\begin{figure}[t]
\centering
\includegraphics[width=0.4\columnwidth]{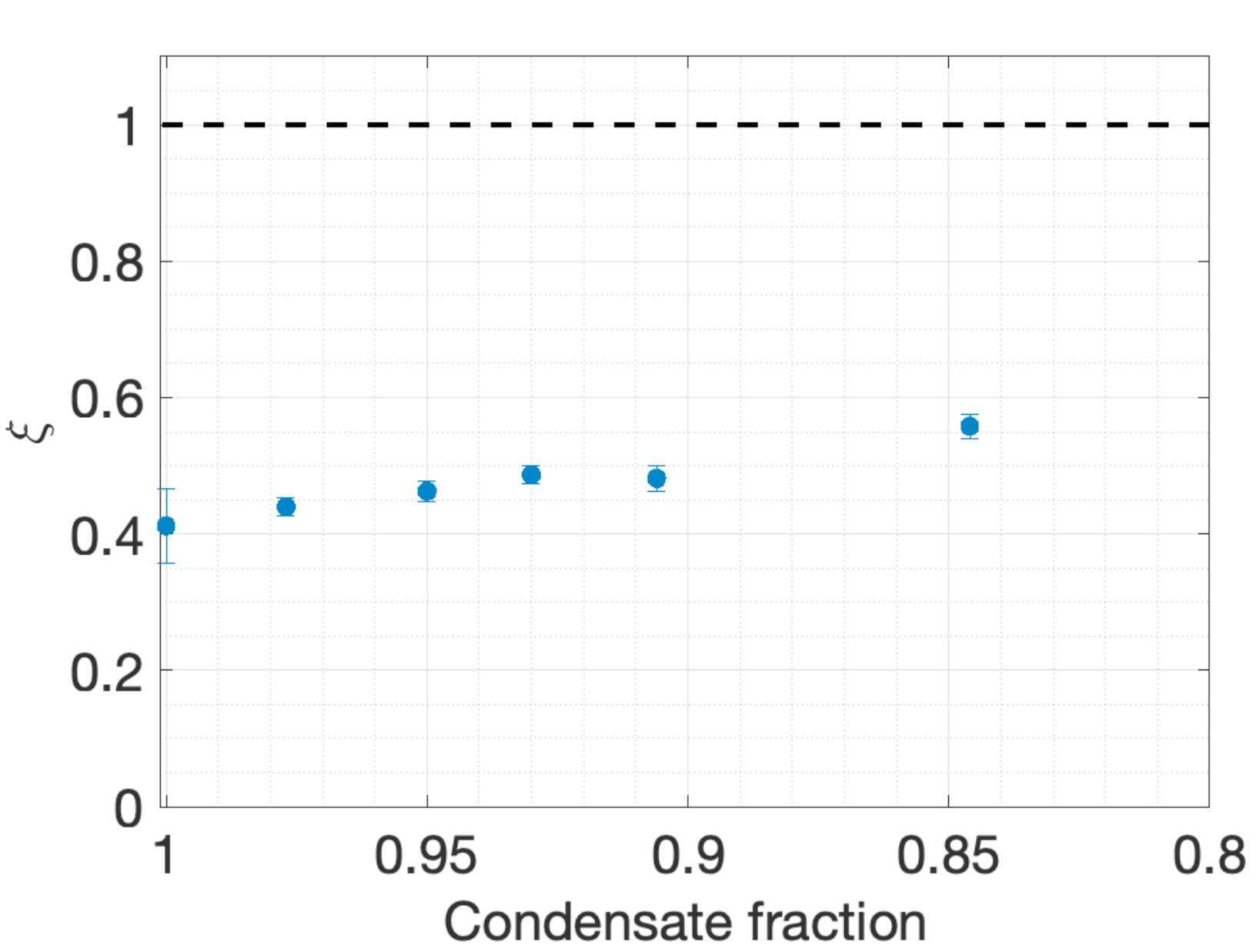}
\caption{Minimum spin squeezing parameter as a function of condensate fraction for $T_\text{OAT} = 10\textrm{ms}$, computed via effective finite temperature 1D TW simulations for an ultracold atomic gas of total atom number $N = 10^4$ in an initial cylindrically-symmetric harmonic trap ($f_r = 32$Hz and $f_z = 160$Hz). The dashed horizontal line indicates the shot-noise limit.
}
\label{fig_finite_temp}
\end{figure}

\section{Optimal beamsplitting parameters for Fig.~3}
Figure~\ref{fig_optimal_BS} shows the optimal second beamsplitting parameters, $\theta$ and $\phi$, that result in the minimum spin squeezing parameters reported in Fig.~3 of the main text. For our analytic model, with model parameters $\lambda$ and $\mathcal{Q} = |\mathcal{Q}| \exp(i \varphi)$ determined from 3D GPE simulations, the optimal beamsplitting angle and phase are given by Eq.~(\ref{theta_opt_sq}) and $\phi_\text{opt} = -\varphi$, respectively. For our TW simulations, the optimal beamsplitting phase was determined via $\phi_\text{opt} = \text{atan2}(-\langle \hat{J}_y^\text{OAT}\rangle, \langle \hat{J}_x^\text{OAT}\rangle)$ (recall that the superscript `OAT' implies that expectations are taken at time $t = 2T_\text{OAT}$ immediately before the second beamsplitter). This maximizes the average pseudospin length $\langle \hat{J}_{\frac{\pi}{2},\phi+\frac{\pi}{2}}\rangle$ by aligning the average pseudopin vector along the $J_x$-axis [see Eq.~(\ref{pseudospin_length})]. $\theta_\text{opt}$ is determined by simply plotting out $\xi_{\theta,\phi_\text{opt}}$ as a function of $\theta$, as shown in Fig.~\ref{fig_scan_theta}, and selecting the minimum.

Although these theoretical estimates can be used to guide experiment, in practice an accurate experimental determination of $\theta$ and $\phi$ would be done by measuring the distribution in the population difference ($\hat{J}_z$) as $\theta$ and $\phi$ are scanned. Specifically, the optimum $\phi$ is chosen by finding the zero-crossing of the interference fringe, as is routinely done in atom interferometry experiments. Once this optimum $\phi$ is determined and fixed, the optimum $\theta$ is found by selecting the $\theta$ that gives the best relative number squeezing after the second beamsplitting pulse. These parameters would then be adjusted to give this state rotated by $\pi/2$ about the $J_x$-axis, i.e. a maximally phase sensitive state at $t = 2T_\text{OAT}$ after the second beamsplitter, as shown in Fig.~1(c) of the main text.

State-of-the-art ultracold-atom gravimetry experiments have demonstrated exquisite control over $\theta$ and $\phi$. The parameter $\phi$ is controlled by the relative phase of the two lasers used to implement the Raman beamsplitter. Specifically, $\phi$ is the change in the phase difference of the two lasers relative to the phase difference of the initial beamsplitter pulse. This parameter is controlled routinely in atom interferometry experiments by making slight adjustments to the two-photon detuning, and is used to map out the interference fringes (see, for example, Ref.~[5]). 
The beamsplitting parameter $\theta$ is the Rabi pulse-area, which determines the relative fraction of population transferred from $|1\rangle$ to $|2\rangle$ (or vice versa). $\theta$ is chosen by either adjusting the intensity of the lasers or the pulse duration.

\begin{figure}[t]
\centering
\includegraphics[width=0.6\columnwidth]{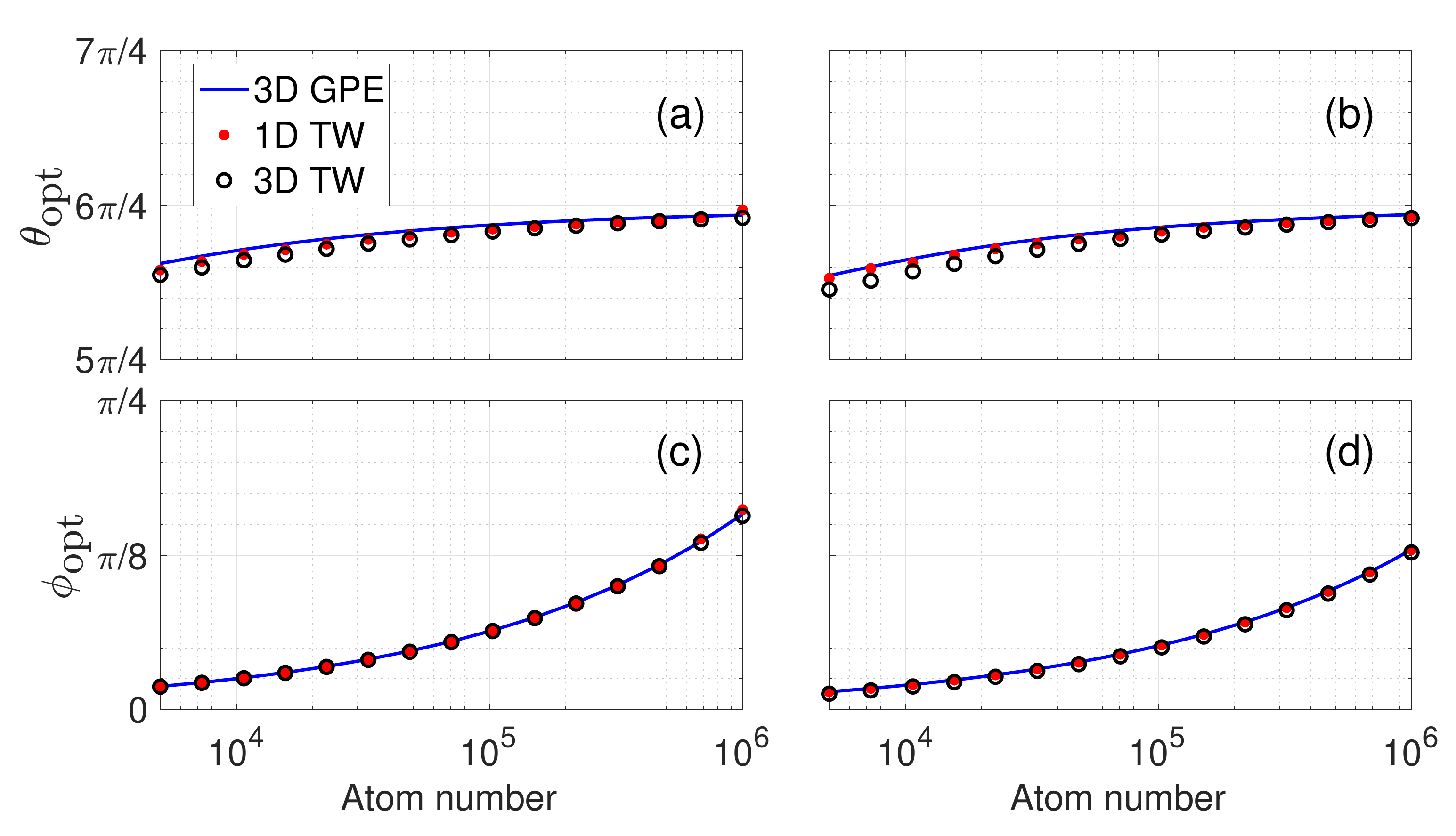}
\caption{Optimal beamsplitting angle (a,b) and phase (c,d) that give the minimum spin squeezing parameters reported in Fig.~3 of the main text. Here the spin squeezing duration is $T_\text{OAT} = 10\textrm{ms}$, the left figures (a,c) correspond to optimal parameters for an initial condensate in a spherically-symmetric harmonic trap ($f_r = f_z = 50$), and the right figures (b,d) correspond to optimal parameters for a `pancake' condensate initially prepared in a cylindrically-symmetric harmonic trap ($f_r = 32$Hz and $f_z = 160$Hz). 
}
\label{fig_optimal_BS}
\end{figure}

\begin{figure}[t]
\centering
\includegraphics[width=0.5\columnwidth]{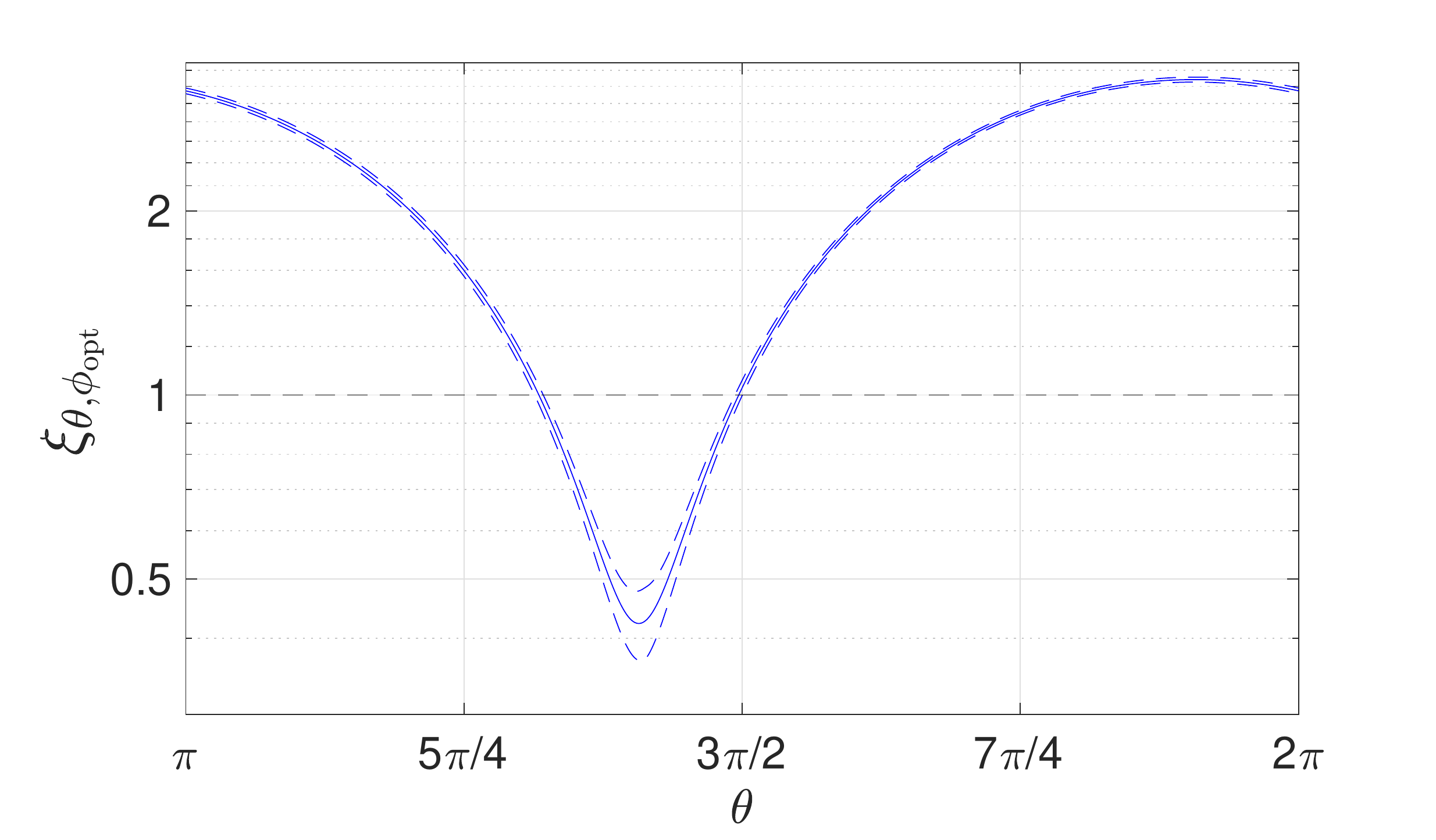}
\caption{Spin squeezing parameter $\xi_{\theta, \phi_\textrm{opt}}$ as a function of 2nd beamsplitting angle $\theta$, computed via 3D TW simulations using Eq.~(\ref{defn_xi}). This is for a spin squeezing duration of $T_\text{OAT} = 10$ms and an initial `pancake' BEC of atom number $N = 10^4$ prepared in a cylindrically-symmetric harmonic potential ($f_r = 32$Hz and $f_z = 160$Hz). Dashed blue lines indicate twice the standard error in the mean (solid line), and the dashed horizontal line indicates the shot-noise limit.
}
\label{fig_scan_theta}
\end{figure}

\section{Effect of shot-to-shot atom number fluctuations}
Here we incorporate shot-to-shot atom number fluctuations into our analytic model (\emph{cf.} above Section `Derivation of analytic model of spin squeezing') and show that the spin squeezing parameter weakly degrades with the size of these fluctuations. We assume that the atom number varies according to a Gaussian distribution
\begin{equation}
	P(N) = \frac{1}{\sqrt{2 \pi \sigma_N^2}}\exp\left[-\frac{(N-N_0)^2}{2 \sigma_N^2}\right],
\end{equation}
where $N_0$ and $\sigma_N^2$ are the mean and variance of the distribution, respectively. Within the two-mode subspace spanned by $\hat{a}_1$ and $\hat{a}_2$, we previously assumed an initial pure state $|N,0\rangle$ (i.e. a $\hat{j}_z$ eigenstate) in order to compute the expectations Eqs.~(\ref{OAT_squeezing_expectations}) (\emph{n.b.} $\hat{a}_1|n_1,n_2\rangle = \sqrt{n_1} |n_1-1,n_2\rangle$ and $\hat{a}_2|n_1,n_2\rangle = \sqrt{n_2} |n_1,n_2-1\rangle$). Here, we instead take our initial state to be the mixture
\begin{equation}
	\hat{\rho} = \sum_{N=0}^\infty P(N) |N,0\rangle \langle N,0|.
\end{equation}
Consequently, the expectation of any operator $\hat{\mathcal{O}}$ is
\begin{equation}
	\langle \hat{\mathcal{O}} \rangle = \int dN P(N) \langle \hat{\mathcal{O}} \rangle_N, \label{ensemble_expectations}
\end{equation}
where $\langle \ldots \rangle_N$ denotes the expectation with respect to an initial Fock state $|N,0\rangle$ and we have taken the continuum limit since $N_0 \gg 1$. In the linear squeezing regime determined by approximations Eqs.~(\ref{eqs_linear_squeezing}), the fixed number expectations directly after OAT are [\emph{cf.} Eqs~(\ref{Big_J_expressions}) and Eqs~(\ref{OAT_squeezing_expectations})]
\begin{subequations}
\begin{align}
	\langle \hat{J}_x^\text{OAT} \rangle_N	&= |\mathcal{Q}_N| \left(  \cos \varphi_N \langle \hat{j}_x^\text{OAT} \rangle_N - \sin \varphi_N \langle \hat{j}_y^\text{OAT} \rangle_N \right) \approx \tfrac{N}{2} |\mathcal{Q}_N| \cos \varphi_N , \\
	\langle \hat{J}_y^\text{OAT} \rangle_N	&= |\mathcal{Q}_N| \left(  \sin \varphi_N \langle \hat{j}_x^\text{OAT} \rangle_N + \cos \varphi_N \langle \hat{j}_y^\text{OAT} \rangle_N \right) \approx \tfrac{N}{2} |\mathcal{Q}_N| \sin \varphi_N , \\
	\langle \hat{J}_z^\text{OAT} \rangle_N	&= \langle \hat{j}_z^\text{OAT} \rangle_N = 0, \\
	\langle (\hat{J}_x^\text{OAT})^2 \rangle_N	&= |\mathcal{Q}_N|^2 \Big( \cos^2 \varphi_N \langle (\hat{j}_x^\text{OAT})^2 \rangle_N + \sin^2 \varphi_N \langle (\hat{j}_y^\text{OAT})^2 \rangle_N \notag \\
									&- \cos \varphi_N \sin \varphi_N \langle \hat{j}_x^\text{OAT} \hat{j}_y^\text{OAT} + \hat{j}_y^\text{OAT} \hat{j}_x^\text{OAT} \rangle_N \Big) + \tfrac{N}{4}\left(1 - |\mathcal{Q}_N|^2\right) \notag \\
							&\approx  \tfrac{N}{4}|\mathcal{Q}_N|^2 \Big( \cos^2 \varphi_N N(1 - N \lambda^2) + \sin^2 \varphi_N (1 + N^2 \lambda^2) \Big) + \tfrac{N}{4}\left(1 - |\mathcal{Q}_N|^2\right), \\
	\langle (\hat{J}_y^\text{OAT})^2 \rangle_N	&= |\mathcal{Q}_N|^2 \Big( \sin^2 \varphi_N \langle (\hat{j}_x^\text{OAT})^2 \rangle_N + \cos^2 \varphi_N \langle (\hat{j}_y^\text{OAT})^2 \rangle_N \notag \\
									&+ \cos \varphi_N \sin \varphi_N \langle \hat{j}_x^\text{OAT} \hat{j}_y^\text{OAT} + \hat{j}_y^\text{OAT} \hat{j}_x^\text{OAT} \rangle_N \Big) + \tfrac{N}{4}\left(1 - |\mathcal{Q}_N|^2\right) \notag \\
							&\approx  \tfrac{N}{4}|\mathcal{Q}_N|^2 \Big( \sin^2 \varphi_N N(1 - N \lambda^2) + \cos^2 \varphi_N (1 + N^2 \lambda^2)\Big) + \tfrac{N}{4}\left(1 - |\mathcal{Q}_N|^2\right), \\
	\langle (\hat{J}_z^\text{OAT})^2 \rangle_N	&= \langle (\hat{j}_z^\text{OAT})^2 \rangle_N = \tfrac{N}{4}, \\
	\tfrac{1}{2}\langle \hat{J}_x^\text{OAT} \hat{J}_y^\text{OAT} + \hat{J}_y^\text{OAT} \hat{J}_x^\text{OAT} \rangle_N	&= \tfrac{1}{2}|\mathcal{Q}_N|^2 \big[ \sin(2 \varphi_N) \big( \langle (\hat{j}_x^\text{OAT})^2 \rangle_N - \langle (\hat{j}_y^\text{OAT})^2 \rangle_N\big) \notag \\
							&+ \cos(2 \varphi_N)\langle \hat{j}_x^\text{OAT} \hat{j}_y^\text{OAT} + \hat{j}_y^\text{OAT} \hat{j}_z^\text{OAT} \rangle_N \big], \notag \\
							&\approx -\tfrac{1}{8} N^3 |\mathcal{Q}_N|^2 \lambda_N^2 \sin(2 \varphi_N), \\
	\tfrac{1}{2}\langle \hat{J}_x^\text{OAT} \hat{J}_z^\text{OAT} + \hat{J}_z^\text{OAT} \hat{J}_x^\text{OAT} \rangle_N	&= |\mathcal{Q}_N| \big( \cos \varphi_N \tfrac{1}{2} \langle \hat{j}_x^\text{OAT} \hat{j}_z^\text{OAT} + \hat{j}_z^\text{OAT} \hat{j}_x^\text{OAT}\rangle_N  - \sin \varphi_N \tfrac{1}{2} \langle \hat{j}_y^\text{OAT} \hat{j}_z^\text{OAT} + \hat{j}_y^\text{OAT} \hat{j}_z^\text{OAT}\rangle_N \big) \notag \\
							&\approx -\tfrac{1}{4}N^2 |\mathcal{Q}_N| \lambda_N \sin \varphi_N, \\
	\tfrac{1}{2}\langle \hat{J}_y^\text{OAT} \hat{J}_z^\text{OAT} + \hat{J}_z^\text{OAT} \hat{J}_y^\text{OAT} \rangle_N	&= |\mathcal{Q}_N| \big( \sin \varphi_N \tfrac{1}{2} \langle \hat{j}_x^\text{OAT} \hat{j}_z^\text{OAT} + \hat{j}_z^\text{OAT} \hat{j}_x^\text{OAT} \rangle_N  + \cos \varphi_N \tfrac{1}{2} \langle \hat{j}_y^\text{OAT} \hat{j}_z^\text{OAT} + \hat{j}_y^\text{OAT} \hat{j}_z^\text{OAT} \rangle_N \big) \notag \\
							&\approx \tfrac{1}{4}N^2 |\mathcal{Q}_N| \lambda_N \cos \varphi_N,
\end{align}
\end{subequations}
where $\lambda_N$ and $\mathcal{Q}_N = |\mathcal{Q}_N| e^{i \varphi_N}$ are the OAT parameter and complex spatial mode overlap [Eq~(\ref{complex_mode_overlap})], respectively, for atom number $N$. For $\sigma_N / N_0 \ll 1$, we do not expect the squeezing strength and mode overlap to substantially vary from shot-to-shot. We therefore approximate $\lambda_N \approx \lambda_{N_0}$, $|\mathcal{Q}_N| \approx |\mathcal{Q}_{N_0}|$, and $\varphi_N \approx \varphi_{N_0}$. Using Eq.~(\ref{ensemble_expectations}), this approximation allows us to derive analytic expressions for the expectations:
\begin{subequations}
\begin{align}
	\langle \hat{J}_x^\text{OAT} \rangle	&= \int dN P(N) \langle \hat{J}_x^\text{OAT} \rangle_N = \tfrac{N_0}{2} |\mathcal{Q}_{N_0}| \cos \varphi_{N_0}, \\
	\langle \hat{J}_y^\text{OAT} \rangle	&= \tfrac{N_0}{2} |\mathcal{Q}_{N_0}| \sin \varphi_{N_0}, \\
	\langle \hat{J}_z^\text{OAT} \rangle	&= 0, \\
	\langle (\hat{J}_x^\text{OAT})^2 \rangle	&= \tfrac{N_0}{8}\left\{ 2 + (N_0-1)|\mathcal{Q}_{N_0}|^2 - |\mathcal{Q}_{N_0}|^2 \left[ 1 - N_0\left( 1 - 2 N_0 \lambda_{N_0}^2 \right) \cos(2 \varphi_{N_0}) \right] \right\} \notag \\
									&+ \tfrac{|\mathcal{Q}_{N_0}|^2}{8}\left[ 1 + \left( 1 - 6 N_0 \lambda_{N_0}^2\right)\cos(2 \varphi_{N_0}) \sigma_N^2 \right], \\
	\langle (\hat{J}_y^\text{OAT})^2 \rangle	&= \tfrac{N_0}{8}\left\{ 2 + (N_0-1)|\mathcal{Q}_{N_0}|^2 + |\mathcal{Q}_{N_0}|^2 \left[ 1 - N_0\left( 1 - 2 N_0 \lambda_{N_0}^2 \right) \cos(2 \varphi_{N_0}) \right] \right\} \notag \\
									&+ \tfrac{|\mathcal{Q}_{N_0}|^2}{8}\left[ 1 - \left( 1 - 6 N_0 \lambda_{N_0}^2\right)\cos(2 \varphi_{N_0}) \sigma_N^2 \right],
\end{align}
\begin{align}
	\langle (\hat{J}_z^\text{OAT})^2 \rangle	&= \tfrac{N_0}{4}, \\
	\tfrac{1}{2}\langle \hat{J}_x^\text{OAT} \hat{J}_y^\text{OAT} + \hat{J}_y^\text{OAT} \hat{J}_x^\text{OAT} \rangle	&= -\tfrac{N_0}{4} |\mathcal{Q}_{N_0}|^2 \left( N_0 + 3 \sigma_N^2\right) \lambda_{N_0}^2 \cos \varphi_{N_0} \sin \varphi_{N_0}, \\
	\tfrac{1}{2}\langle \hat{J}_x^\text{OAT} \hat{J}_z^\text{OAT} + \hat{J}_z^\text{OAT} \hat{J}_x^\text{OAT} \rangle	&= -\tfrac{N_0}{4} |\mathcal{Q}_{N_0}| \left( N_0^2 + \sigma_N^2\right) \lambda_{N_0} \sin \varphi_{N_0}, \\
	\tfrac{1}{2}\langle \hat{J}_y^\text{OAT} \hat{J}_z^\text{OAT} + \hat{J}_z^\text{OAT} \hat{J}_y^\text{OAT} \rangle	&= \tfrac{N_0}{4} |\mathcal{Q}_{N_0}| \left( N_0^2 + \sigma_N^2\right) \lambda_{N_0} \cos \varphi_{N_0}.
\end{align}
\end{subequations}
This gives
\begin{equation}
	\xi_{\theta,\varphi}^2(\tfrac{\sigma_N}{N_0}) = \xi_{\theta,\varphi}^2 + \tfrac{N_0 \lambda_{N_0}}{ |\mathcal{Q}_{N_0}|}\sin \theta \left[ 3 N_0  |\mathcal{Q}_{N_0}| \lambda_{N_0} \sin \theta - 2 \cos \theta \sec \left( \varphi + \varphi_{N_0}\right)\right] \left(\tfrac{\sigma_N}{N_0}\right)^2,
\end{equation}
where $\xi_{\theta,\varphi}$ is the spin squeezing parameter in the linear squeezing regime in the limit of zero shot-to-shot atom number fluctuations [see Eq.~(\ref{general_spin_squeezing_parameter})]:
\begin{equation}
	\xi_{\theta,\varphi}^2 = N_0 (1 + N_0 \lambda_{N_0}^2) \sin^2 \theta + \frac{\sec(\phi + \varphi_{N_0})}{|\mathcal{Q}_{N_0}|^2} \left[ \sec(\phi + \varphi_{N_0})\left( \cos^2 \theta + (1 - N_0 |\mathcal{Q}_{N_0}|^2) \sin^2 \theta \right) - N_0 |\mathcal{Q}_{N_0}| \lambda_{N_0} \sin(2 \theta) \right].
\end{equation}
In the $\sigma_N \to 0$ limit, the spin squeezing parameter is minimized for the choice $\theta = \theta_\text{sq} = \tfrac{3\pi}{2} - \tfrac{1}{2}\tan^{-1}[2/(N_0 |\mathcal{Q}_{N_0}| \lambda_{N_0})]$ [Eq.~(\ref{theta_opt_sq})] and $\phi = - \varphi_{N_0}$, yielding
\begin{equation}
	\xi^2(\tfrac{\sigma_N}{N_0}) = \xi^2 + \left[\frac{3}{2} N_0^2 \lambda_{N_0}^2 |\mathcal{Q}_{N_0}|^2 \left(\frac{\sqrt{4 + N_0^2  |\mathcal{Q}_{N_0}|^2 \lambda_{N_0}^2} -N_0  |\mathcal{Q}_{N_0}| \lambda_{N_0} - \frac{4}{3 N_0  |\mathcal{Q}_{N_0}| \lambda_{N_0}} }{\sqrt{4 + N_0^2  |\mathcal{Q}_{N_0}|^2 \lambda_{N_0}^2}}\right)\right] \frac{1}{|\mathcal{Q}_{N_0}|^2}\left(\frac{\sigma_N}{N_0}\right)^2,
\end{equation} 
where $\xi = \xi_{\theta_\text{sq}, -\varphi_{N_0}}$ is given by Eq.~(3) of the main text. Since $N_0 |\mathcal{Q}_{N_0}| \lambda_{N_0} \geq 0$, the term in square brackets is bounded from above by 1. Therefore
\begin{equation}
	\xi(\tfrac{\sigma_N}{N_0}) \leq \sqrt{\xi^2 + \tfrac{1}{|\mathcal{Q}|^2}\left(\tfrac{\sigma_N}{N_0}\right)^2} \approx \xi + \tfrac{1}{2|\mathcal{Q}_{N_0}|^2}\left(\tfrac{\sigma_N}{N_0}\right)^2,
\end{equation}
as reported in the main text.

\end{widetext}

\end{document}